\documentclass[fleqn,usenatbib]{mnras}

\usepackage{newtxtext,newtxmath,amsmath,siunitx}
\usepackage[T1]{fontenc}

\DeclareRobustCommand{\VAN}[3]{#2}
\let\VANthebibliography\thebibliography
\def\thebibliography{\DeclareRobustCommand{\VAN}[3]{##3}\VANthebibliography}

\usepackage{graphicx}	% Including figure files
\usepackage{amsmath}	% Advanced maths commands
\usepackage{longtable}
\usepackage[normalem]{ulem}

\newcommand{\frb}{FRB~20121102A}

\title[\frb\ burst storm in 2016]{Arecibo observations of a burst storm from \frb\ in 2016}

\author[D. M. Hewitt et al.]{
D.~M.~Hewitt,$^{1}$\thanks{E-mail: d.m.hewitt@uva.nl}
M.~P.~Snelders,$^{1}$
J.~W.~T.~Hessels,$^{1,2}$
K.~Nimmo,$^{2,1}$
J.~N.~Jahns,$^{3}$
L.~G.~Spitler,$^{3}$
\newauthor K.~Gourdji,$^{1}$
G.~H.~Hilmarsson,${^3}$
D.~Michilli,${^{4,5}}$
O.~S.~Ould-Boukattine,$^{1}$
P.~Scholz,$^{6}$ A.D.~Seymour$^{7}$
\\
$^{1}$Anton Pannekoek Institute for Astronomy, University of Amsterdam, Science Park 904, 1098 XH Amsterdam, The Netherlands\\
$^{2}$ASTRON, Netherlands Institute for Radio Astronomy, Oude Hoogeveensedijk 4, 7991 PD Dwingeloo, The Netherlands\\
$^{3}$Max-Planck Institute for Radio Astronomy, Auf dem Hügel 69, D-53121 Bonn, Germany\\
$^{4}$MIT Kavli Institute for Astrophysics and Space Research, Massachusetts Institute of Technology, 77 Massachusetts Ave, Cambridge, MA 02139, USA\\
$^{5}$Department of Physics, Massachusetts Institute of Technology, 77 Massachusetts Ave, Cambridge, MA 02139, USA\\
$^{6}$Dunlap Institute for Astronomy \& Astrophysics, University of Toronto, 50 St. George Street, Toronto, ON M5S 3H4,
Canada\\
$^{7}$Green Bank Observatory, P.O. Box 2, WV 24944, Green Bank, USA \\
}

\date{Accepted XXX. Received YYY; in original form ZZZ}

\pubyear{2021}

\begin{document}
\label{firstpage}
\pagerange{\pageref{firstpage}--\pageref{lastpage}}
\maketitle

\begin{abstract}
\frb\ is the first known fast radio burst (FRB) from which repeat bursts were detected, and one of the best-studied FRB sources in the literature. Here we report on the analysis of 478 bursts (333 previously unreported) from \frb\ using the 305-m Arecibo telescope --- detected during approximately 59 hours of observations between December 2015 and October 2016. The majority of bursts are from a burst storm around September 2016.  This is the earliest available sample of a large number of \frb\ bursts, and it thus provides an anchor point for long-term studies of the source's evolving properties. We observe that the bursts separate into two groups in the width-bandwidth-energy parameter space, which we refer to as the low-energy bursts (LEBs) and high-energy bursts (HEBs). The LEBs are typically longer duration and narrower bandwidth than the HEBs, reminiscent of the spectro-temporal differences observed between the bursts of repeating and non-repeating FRBs.  We fit the cumulative burst rate-energy distribution with a broken power-law and find that it flattens out toward higher energies. The sample shows a diverse zoo of burst morphologies.  Notably, burst emission seems to be more common at the top than the bottom of our $1150-1730$\,MHz observing band.  We also observe that bursts from the same day appear to be more similar to each other than to those of other days, but this observation requires confirmation. The wait times and burst rates that we measure are consistent with previous studies.  We discuss these results, primarily in the context of magnetar models.   
\end{abstract}


%%%%%%%%%%%%%%%%%%%%%%%%%%%%%%%%%%%%%%%%%%%%%%%%%%

%%%%%%%%%%%%%%%%% BODY OF PAPER %%%%%%%%%%%%%%%%%%

\section{Introduction}

% Introduction to FRBs
Fast radio bursts (FRBs) are flashes of radio emission with exceptionally high isotropic equivalent luminosity that usually last no longer than a few tens of milliseconds and originate at extragalactic distances \citep{lorimer_2007_sci,thornton_2013_sci,petroff_2019_aarv,petroff_2021_arxiv,cordes_2019_araa}. Hundreds of FRBs have been detected to date \citep[see the Transient Name Server\footnote{\url{https://www.wis-tns.org}} or][]{petroff_2016_pasa,chime_2021_arxiv_arxiv210604352}, but only approximately 20 of these sources are known to repeat \citep[e.g.,][]{spitler_2016_natur,shannon_2018_natur,kumar_2019_apjl,chime_2019_natur_566_235,chime_2019_apjl,fonseca_2020_apjl}. Over a dozen FRBs (both repeating as well as apparently non-repeating) have been localised to a variety of host galaxies \citep[][]{bhandari_2021_arxiv}, but the emission mechanism and physical origin remain enigmatic. While the repeaters rule out cataclysmic models for at least some fraction of the FRB population, the models that have been proposed for FRBs span a wide range of astrophysical scenarios \citep[for a catalogue of models see][]{platts_2019_phr}. Recently, there has been the noteworthy detection of an extremely bright radio burst from the Galactic magnetar SGR~J1935+2154 \citep{bochenek_2020_natur,chime_2020_natur_587}, which has strengthened the case for magnetars as the engines of FRBs.  Even so, the discovery of a nearby repeating FRB in a globular cluster shows that not all FRBs can be young magnetars formed via core collapse \citep[][]{kirsten_2021_arxiv} --- though they may be, in some cases, young magnetars formed via accretion-induced collapse or binary merger.

% Repeaters vs. non-repeaters
It remains unclear whether the repeating and apparently non-repeating FRBs come from physically distinct source types. Recently, it has been shown that the burst properties of repeating and non-repeating FRBs are statistically different, with repeaters on average exhibiting larger temporal widths and narrower bandwidths \citep{pleunis_2021_arxiv, chime_2021_arxiv_arxiv210604352}. Additionally, many repeater bursts display  complex time-frequency downward-drifting structure \citep[the `sad-trombone' effect;][]{hessels_2019_apjl}. Another commonality that seems to be emerging among repeating sources, is near 100 percent linear polarisation, a very low circular polarisation fraction and polarisation position angles (PPAs) that remain approximately constant between and during bursts \citep{michilli_2018_natur,nimmo_2021_natas}. While there have been a small number of detections of circular polarisation in repeater bursts \citep{hilmarsson_2021_mnras,kumar_2021_arxiv} and swinging PPAs \citep[][]{luo_2020_natur}, the non-repeating FRB population shows more diverse polarimetric properties \citep[e.g.,][]{day_2020_mnras}. These differences in the burst properties of repeaters and apparent non-repeaters could indicate different emission mechanisms, different progenitors and/or different local environments.

% Introduce FRB 121102
\frb\ is the first FRB source from which multiple bursts were detected \citep{spitler_2014_apj,spitler_2016_natur}. This repeating nature allowed for follow-up observations that localised the source to near a star-forming region in a dwarf galaxy at redshift $z=0.193$ \citep{bassa_2017_apjl,chatterjee_2017_natur,tendulkar_2017_apjl} and, moreover, a milliarcsecond association of the bursts with a persistent compact radio source \citep{marcote_2017_apjl,chatterjee_2017_natur}. The most recent dispersion measure (DM) values of the bursts from \frb\ are $563.5\pm0.8$\,pc\,cm$^{-3}$ on MJD~57836 \citep[][]{platts_2021_mnras} and  $565.8\pm0.8$\,pc\,cm$^{-3}$ for the period between MJD~58724 and MJD~58776 \citep[][]{li_2021_natur}. \citet[][]{li_2021_natur} also estimated that this DM is increasing by approximately $1$\,pc\,cm$^{-3}$\,yr$^{-1}$. Moreover, there is evidence that this secular DM increase is non-linear \citep[][; Seymour et al. \textit{in prep.}]{Jahns_2022_arXiv}.  \cite{michilli_2018_natur} showed that \frb\ has an extremely large and variable rotation measure (RM $\sim 10^5\,$rad\,m$^{-2}$), demonstrating that the bursts originate from a source embedded in an extreme and dynamic magneto-ionic environment. In contrast to the DM, the RM of the bursts is decreasing by an average of $\sim15$\,percent per year \citep{hilmarsson_2021_apjl}.  \citet{Plavin_2022_arXiv} showed that the source depolarises towards lower frequencies, possibly because of minor non-uniformities in the Faraday screen (they find that the Faraday width
of the burst environment is only approximately 0.1 per cent of the total Faraday rotation).

\frb\ is in many ways the prototypical repeating FRB.  As with other repeaters, the burst profile and spectrum vary significantly from burst to burst. Furthermore, the burst activity of \frb\ appears to be frequency dependent, an aspect that has not been well studied in most other sources \citep[though see, e.g.,][]{pleunis_2021_apjl}. While thousands of bursts were detected in less than two months in the $1-2$\,GHz range \citep[][]{li_2021_natur}, only one burst has been detected in the $400-800$\,MHz range  \citep[more specifically at $\sim600$\,MHz;][]{josephy_2019_apjl} despite years of nearly daily monitoring by the Canadian Hydrogen Intensity Mapping Experiment Fast Radio Burst Project (CHIME/FRB). On the other extreme, the highest frequency at which bursts have been detected from \frb\ (and from any FRB for that matter) is $\sim$8\,GHz \citep{gajjar_2018_apj}. From a temporal point of view, the bursts from \frb\ are clustered in time \citep{oppermann_2018_mnras}, but show no evidence of a rotational period underlying their times-of-arrival \citep{zhang_2018_apj,aggarwal_2021_arxiv}. Analysis of burst data over 5 years has, however, resulted in an activity period of $\sim160$ days and a duty cycle of $\sim55$\,percent \citep[][]{rajwade_2020_mnras,cruces_2021_mnras}. \frb\ is one of only two known FRB sources with a long-term periodicity: the other being FRB~20180916B, which has a well-established activity period of 16.35 days with an activity window of $\sim$5 days. \citep[][]{chime_2020_natur_582}.

%repeater models
Various models have been proposed to explain the repeating nature and periodic activity of some FRBs, typically by either invoking  precession of a neutron star \citep[NS; e.g.,][]{Levin_2020_ApJL,sobyanin_2020_mnras,Yang_2020_ApJL,zanazzi_2020_apjl}, the rotational period of a NS \citep[e.g.,][although \textit{such} slow-rotating magnetars have not yet been observed]{beniamini_2020_mnras_496} or the orbital period of a NS in a binary system with another astrophysical object \citep[e.g.,][]{gu_2020_mnras,lyutikov_2020_apjl,du_2021_mnras,kuerban_2021_arxiv,sridhar_2021_apj,wada_2021_apj}.
The literature is also divided on the location of the emission region, assuming a NS origin: i.e., whether this is within the magnetosphere of the NS \citep[e.g.,][]{Kumar_2017_MNRAS}, or much further out beyond the magnetosphere as the result of relativistic shocks \citep[e.g.,][]{Metzger_2019_MNRAS}. 

%what we can do to differentiate between these models
The spectro-temporal and polarimetric properties of the bursts, as well as their energetics and the evolution of these factors, are all valuable probes for constraining and differentiating between progenitor models. 
The short time intervals between bursts, as well as the high burst rate detected by FAST (Five hundred meter Aperture Spherical Telescope) favour those models where burst genesis occurs in the magnetosphere \citep[][]{li_2021_natur}.  Burst structure down to $\sim4\,\mu$s and $\sim60$\,ns  is detected in FRB~20180916B and FRB~20200120E, respectively \citep{nimmo_2021_arxiv,nimmo_2021_natas}. Ignoring relativistic effects, these durations correspond to an emission region from $\sim100$\,m to a few km in size, which arguably supports magnetospheric models \citep[][]{beniamini_2020_mnras_498}. The sad-trombone effect, on the other hand can be explained as a radius-to-frequency mapping effect in magnetospheric models \citep[][]{lyutikov_2020_APJ}, but is also a natural consequence of models invoking a synchrotron maser \citep[e.g.,][]{Metzger_2019_MNRAS,margalit_2020_apjl}.

% Outline of paper
Through a monitoring proposal (P3054; PI: L. Spitler), the Arecibo Observatory's 305-m telescope has been used to detect over a thousand bursts from \frb\ since its discovery \citep[see also][]{Jahns_2022_arXiv}.  In this paper we present analysis of the earliest observed burst storm of \frb. In \S\ref{sec:observations}, we describe the observations and our search analysis used to discover the bursts presented in this work. In \S\ref{sec:results} we present the search results and explain how the burst properties have been measured. We describe how we classify bursts as either `low-energy bursts' or `high-energy bursts', and evaluate the statistical significance of this classification. We also explain our procedures of fitting power-laws to the energy distribution of the detected bursts. In \S\ref{sec:discussion} we investigate how the classification of low- and high-energy bursts relates to what is known from other FRBs. Furthermore, we discuss the results of the energy and spectro-temporal properties of this sample of bursts. Finally, in \S\ref{sec:conclusions} we summarise our main findings.

\section{Observations and Search for Bursts} \label{sec:observations}

We observed \frb\ with the 305-m William E. Gordon Telescope at the Arecibo Observatory using the L-wide receiver. In this paper, we report on 56 observations that sum to $\sim59\,$hours on source. These observations were conducted on different days, sometimes consecutive, between November 2015 and October 2016, and are summarised in Table~\ref{tab:observations}. The data were recorded using the Puerto-Rican Ultimate Pulsar Processing Instrument (PUPPI) back-end recorder\footnote{\url{http://www.naic.edu/puppi-observing/}}, which records from $980-1780$\,MHz and provides a time resolution of 10.24\,$\mu$s and 512 frequency channels of width 1.56\,MHz. The L-wide receiver has a smaller observing frequency range of 1150$-$1730\,MHz, and the data outside this frequency range are excluded from analysis. These data were coherently dedispersed \textit{within} a channel to a DM of 557.0\,pc\,cm$^{-3}$, the best DM measurement at the time of the observations, which effectively reduces intra-channel dispersive smearing to $<30\,\mu$s given a true DM of 560.5\,pc\,cm$^{-3}$ \citep[][]{hessels_2019_apjl}.

Prior to searching for bursts, we downsampled the data using \texttt{psrfits\_subband}\footnote{\url{https://github.com/demorest/psrfits\_utils}}, resulting in 64 channels of 12.5\,MHz and a time resolution of 81.92\,$\mu$s. Using \texttt{PRESTO}\footnote{\url{https://www.cv.nrao.edu/~sransom/presto/}} \citep{ransom_2001_phdt}, the data were searched for bursts by first incoherently dedispersing the time series using \texttt{prepsubband}, for 77 equally spaced trial DMs in the range $554.90-570.10$\,pc\,cm$^{-3}$ (steps of $0.2$\,pc\,cm$^{-3}$). No radio frequency interference (RFI) masks were applied, in order to prevent masking very bright bursts. After this, the de-dedispersed time series were searched with \texttt{single\_pulse\_search.py}, which implements matched boxcar filtering. Our search was most sensitive to bursts with boxcar widths between 81.92\,$\mu$s and 24.58\,ms.  We searched down to events with a S/N of 6, as provided by \texttt{single\_pulse\_search.py}. The detected events were grouped into burst candidates using \texttt{SpS}\footnote{\url{https://github.com/danielemichilli/SpS}} \citep{michilli_2018_mnras}. Events within 20\,ms and 1 \,pc\,cm$^{-3}$ are grouped together at first. This chosen time window for grouping is motivated by the typical burst duration.  Since each grouped burst is manually inspected over a larger time range, it is then possible to identify these bursts as multiple bursts, and/or to find additional weaker bursts that are closely spaced in time.

The candidates were then classified with \texttt{FETCH}\footnote{\url{https://github.com/devanshkv/fetch}} \citep[Fast Extragalactic Transient Candidate Hunter;][]{agarwal_2020_mnras}, a supervised deep-learning algorithm that uses deep neural networks to classify burst candidates. \texttt{FETCH} has been trained on simulated FRBs and real RFI from the Green Bank Observatory's (GBT) L-band data (comparable frequencies to the Arecibo data presented in this paper). There are eleven models that all perform well on GBT data, and have also proven very successful at detecting FRBs with S/N$>$10 in ASKAP (Australian Square Kilometre Array Pathfinder) and Parkes data. Since the Arecibo Observatory has additional disparate sources of RFI, and since we also probe to a lower S/N than the training data, we first evaluated the performance of the different models through manual inspection using previously searched datasets which contained 41 known bursts \citep{gourdji_2019_apjl}. We found that model H performed the best at identifying true bursts. We managed to recover all bursts except for B33 (which is missed by all available models) from \citet[][]{gourdji_2019_apjl}. In \citet[][]{Jahns_2022_arXiv}, where their data are very similar\footnote{The same telescope, receiver, backend and source but a factor $8$ higher frequency resolution.} to those presented in this paper, they calculated the completeness of FETCH. When compared to a completely human-labeled classification, they also found that model H misses the least amount of bursts. In their analysis they found that $90$\% of the bursts that are missed have a S/N $<$ 8 --- the minimum S/N that \texttt{FETCH} has been trained on. They also find that, in their case, $34$\% of the bursts with S/N $<$ 8 deemed real (by eye) score a probability $p<0.5$. They note that \texttt{FETCH} manages to recover $96$\% of the bursts with S/N $\geqslant$ 8 and mention that the ones that are missed are strongly affected by (broadband) RFI. All diagnostic plots of our candidates to which model H assigned a probability $p\geqslant0.5$ had to undergo two independent by-eye inspections to be deemed true astrophysical bursts.

Two out of our 56 datasets have previously been analysed by both \citet[][]{gourdji_2019_apjl} and \citet[][]{aggarwal_2021_arxiv}. \citet[][]{aggarwal_2021_arxiv} also miss B33 from \citet[][]{gourdji_2019_apjl}, and speculate that it may be due to the narrowbandedness and low S/N of the burst. Additionally, \citet[][]{hessels_2019_apjl} have reported on 14 bright bursts that are also included in our analysis here. We emphasise that even despite the large number of bursts found, there are possibly still a small number of bursts being missed due to the range of our boxcar widths used in searching, S/N-limitations and the model predictions of \texttt{FETCH}. Our sensitivity to bursts some factor shorter or longer than our minimum and maximum boxcar lengths decreases as a function of the square root of said factor. Furthermore, we do not implement any subbanded searches, which means we are less sensitive to very narrowband bursts from \frb\ (bursts that extend over less than approximately a third of the observing bandwidth). 

\begin{table}
\centering
    \caption{Observation log}
    \label{tab:observations}
    \begin{tabular}{llrr} \hline \hline
    Topocentric Start MJD            & Duration (s)  & Nr of Bursts & Burst Rate (hr$^{-1}$)\\ \hline
    
57342.225104167 & 5753.9 & 0 & 0 \\
57343.223344907 & 6747.7 & 0 & 0 \\
57344.224456019 & 6421.4 & 0 & 0 \\
57345.227222222 & 3000.1 & 0 & 0 \\
57345.263599537 & 1300.3 & 0 & 0 \\
57345.280393519 &  971.9 & 0 & 0 \\
57352.220868056 & 2705.5 & 0 & 0 \\
57364.196805556 &  245.4 & 1 & 14.7 \\
57381.179918981 & 1534.4 & 0 & 0 \\
57388.099490741 & 5858.8 & 0 & 0 \\
57391.152835648 & 1200.8 & 0 & 0 \\
57391.131307870 & 1700.2 & 0 & 0 \\
57410.039421296 & 6485.3 & 0 & 0 \\
57413.040474537 & 6058.0 & 0 & 0 \\
57418.019942130 & 6362.4 & 0 & 0 \\
57420.016736111 &  465.2 & 0 & 0 \\
57420.023229167 &  524.2 & 0 & 0 \\
57420.029490741 &  606.5 & 0 & 0 \\
57420.036678241 &  605.9 & 0 & 0 \\
57420.043888889 &  598.4 & 0 & 0 \\
57420.050983796 &  603.2 & 0 & 0 \\
57420.058148148 &  602.0 & 0 & 0 \\
57420.065289352 &  619.2 & 0 & 0 \\
57428.986134259 & 6774.8 & 0 & 0 \\
57429.988379630 & 6487.5 & 0 & 0 \\
57504.819814815 &  625.5 & 0 & 0 \\
57504.833935185 & 2175.1 & 0 & 0 \\
57506.804212963 & 4262.7 & 1 & 0.8\\
57510.801192130 & 3579.1 & 4 & 4.0\\
57511.836574074 &  292.8 & 0 & 0 \\
57512.759571759 & 6709.4 & 0 & 0 \\
57513.786701389 & 4125.2 & 1 & 0.87 \\
57514.809259259 & 1945.5 & 0 & 0 \\
57523.730046296 & 6659.5 & 0 & 0 \\
57571.597083333 & 6831.0 & 0 & 0 \\
57578.579027778 & 6661.3 & 0 & 0 \\
57586.561643519 & 6351.8 & 0 & 0 \\
57594.541250000 & 6232.9 & 0 & 0 \\
57607.507083333 & 6326.2 & 10 & 5.7\\
57614.479687500 & 7179.5 & 48 & 24.1\\
57621.460729167 & 6818.5 & 0 & 0 \\
57628.446423611 & 3615.3 & 44 & 43.8\\
57638.448877315 & 3831.5 & 39 & 36.6\\
57640.412835648 &  808.3 & 10 & 44.5\\
57640.435451389 &  307.4 & 1 & 11.7\\
57640.464143519 & 2277.3 & 10 & 15.8\\
57641.438032407 & 4059.9 & 34 & 30.1\\
57642.439664352 & 3683.7 & 8 & 7.8\\
57644.407719907 & 5967.1 & 56 & 33.8\\
57645.409861111 & 5545.3 & 76 & 49.3\\
57646.394166667 & 3751.5 & 24 & 23.0\\
57648.393437500 & 6036.4 & 29 & 17.3\\
57649.394710648 & 5672.7 & 26 & 16.5\\
57650.380810185 & 6595.3 & 15 & 8.2\\
57666.338912037 & 6723.1 & 41 & 22.0\\
57671.374085648 & 2501.7 & 0 & 0 \\

\textbf{Total} & 212386.1 & 478 \\
\hline
    \end{tabular}
\end{table}

\begin{figure*}
    \includegraphics[width=1\textwidth]{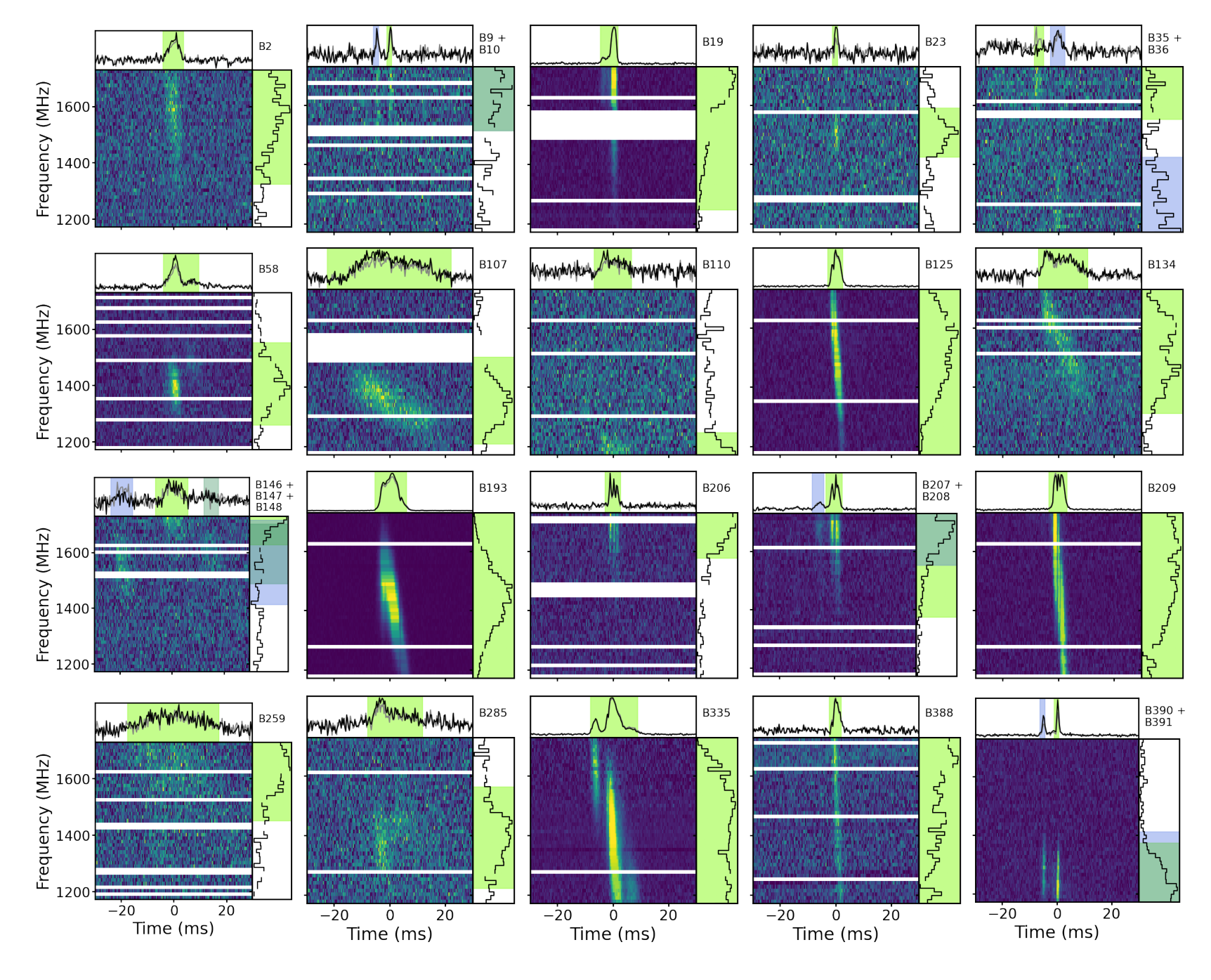}
	\caption{Dynamic spectra of a sub-sample of the 478 detected bursts from \frb. This sample contains bursts that are extremely bright,  exhibit complex morphology or inhabit the extremes of the observing frequency range. Table~\ref{tab:allbursts} is a sample of an online table which contains all 478 bursts and their properties. All bursts have been dedispersed to a DM of 560.5 pc\,cm$^{-3}$. The horizontal white stripes are channels that have been excised in order to remove radio frequency interference. The top panel of each thumbnail shows two burst profiles: one from summing over the entire observing band (in grey) and a second from only summing over the spectral extent where the burst is detected (in black). The panels on the right show the on-pulse time-averaged spectra. The green shaded areas indicate the 2$\sigma$ region of Gaussian fits to the burst profile and spectrum. In the case were multiple bursts are present within a few tens of milliseconds, blue and darker green shaded areas indicate the 2$\sigma$ region for these additional bursts. The burst number is displayed in the top right of each thumbnail. The dynamic spectra have been downsampled to a time and frequency resolution of 327.68\,$\rm{\mu}$s and 12.5\,MHz, respectively. For visual purposes, the dynamic range of the colour scale is adjusted to saturate below the fifth percentile and above 95 percent of the peak signal. Frequency and time axes are scaled equally to highlight the diversity in burst morphology. Three of the bursts displayed above were also presented in \citet[][]{hessels_2019_apjl} at higher time resolution with `AO' names: B125 (AO-02), B193 (AO-05), B209 (AO-06).}
    \label{fig:dynamic_spectra}
\end{figure*}

\section{Results and Analysis} \label{sec:results}

\begin{figure*}
    \centering
    \includegraphics[width=0.85\textwidth]{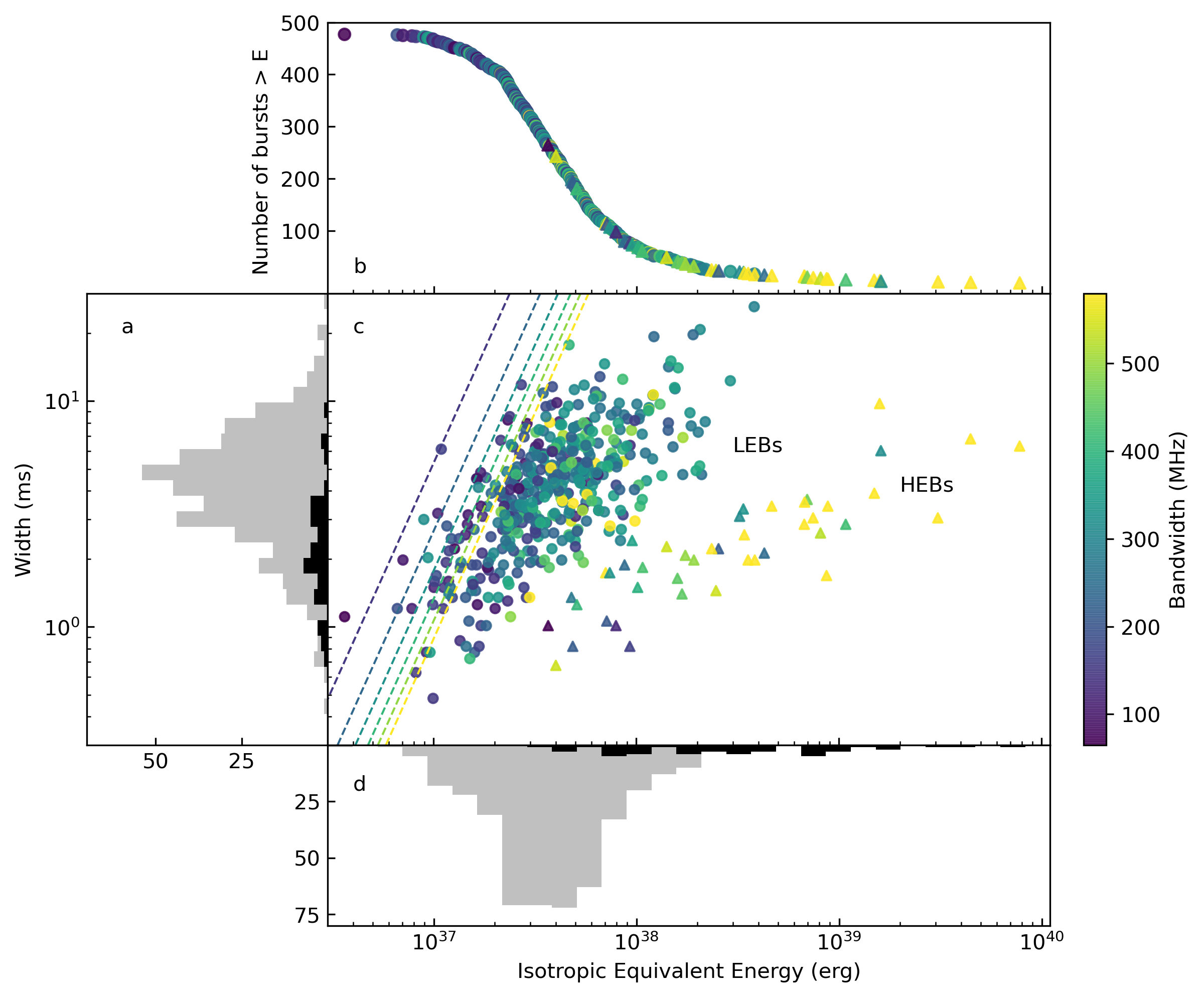}

	\caption{The 478 bursts have been clustered into two groups based on their width and energy as defined in \S\ref{sec:two_groups}, and these two clusters are depicted here as circles (the low-energy bursts, or LEBs) and triangles (the high-energy bursts, or HEBs). \textbf{a:} The width (FWHM) distributions of the bursts. The grey and black histograms correspond to the bursts indicated by circles (LEBs) and triangles (HEBs), respectively. \textbf{b:} The cumulative burst number-energy distribution. The data points are coloured according to the observed bandwidth of the bursts and correspond to the data points shown in Panel c. \textbf{c:} The burst width plotted against the isotropic equivalent energy for all bursts.  The detection thresholds for bursts with a S/N of 6 and various bandwidths (multiples of 100\,MHz) are indicated by slanted dashed lines, which are also coloured according to observed bandwidth using the same scale as for the data points.  \textbf{d:} Same as Panel a, but for isotropic equivalent burst energy rather than temporal width.}
    \label{fig:e_v_w}
\end{figure*}

A total of 478 bursts were determined to be of astrophysical origin in these observations, most of which were low S/N (see Figure~\ref{fig:sn}). The majority (333) of these bursts have not been reported before. Due to the grouping of pulses within 20\,ms into burst candidates, there were a small number of instances ($\sim7$\,percent) where other nearby, much fainter bursts were also visible in addition to those pulses classified by \texttt{FETCH}. These bursts have been included in the total number of 478. Distinguishing between bursts and sub-bursts is non-trivial and becomes ambiguous at a certain level. In order to do so, we thus implemented the following definition: if a distinct Gaussian-like component in the frequency-averaged burst profile does not smear into other components but instead begins and ends at a baseline comparable to the noise, it is defined as a burst, unless the component exists as one of the downward-drifting components often seen from repeaters (the sad-trombone effect), in which case we define it as a sub-burst.  Figure~\ref{fig:dynamic_spectra} shows the dynamic spectra of a sub-sample of bursts that are either bright, short duration or narrowband, or have complex time-frequency structure. We do not see any clear upward drifting bursts (a `happy trombone' effect) in this sample. 

All bursts were dedispersed to a DM of 560.5\,pc\,cm$^{-3}$, determined by \cite{hessels_2019_apjl}, which maximised the band-integrated structure of the bursts and provided a smaller range (more than an order of magnitude) in the inferred DMs of the bursts than the DM that optimises peak S/N. To read in the fits files, we make use of the \texttt{PsrfitsFile.read\_subint} function from the \texttt{psrfits.py} script of \texttt{PRESTO}, which reads in sub-intervals from a fits file. When doing so, the scales, weights and offsets had to be explicitly applied to prevent saturation of the brighter bursts as a result of previous downsampling. Given that the Arecibo observations in \citet[][]{hessels_2019_apjl} are a sub-sample of the observations presented here, we choose this DM and assume that any change in the DM over the course of our observations is negligible. For each burst dynamic spectrum, the channels in which RFI were present were flagged and excluded from further processing. The dynamic spectra were then corrected for bandpass variations by subtracting the off-burst mean and dividing by the off-burst standard deviation, per channel. In Figure~\ref{fig:dynamic_spectra}, in addition to dynamic spectra, we also show a time averaged spectrum, a full-band averaged burst time profile and the burst time profile obtained by only averaging over the frequencies where the burst is present (the 2$\sigma$ region of a one-dimensional Gaussian fit to the burst spectrum). A one-dimensional Gaussian function is also fit to each full-band burst profile. The burst spectra are produced by time-averaging the dynamic spectrum over the 2$\sigma$ region of this Gaussian fit. The 2$\sigma$ regions of both Gaussian fits are indicated by green (and blue) shaded regions. We define the temporal width of a burst as the FWHM of the band-limited burst profile, and bandwidth as the 2$\sigma$ region of the fit to the spectrum. We calculated the fluence of a burst by summing over the 2$\sigma$ time region of a burst and then multiplying each channel of the spectrum with a channel-specific scaling based on the radiometer equation. Thereafter, the mean of the on-pulse spectrum is multiplied with the sampling time. This is equivalent to taking the product of the sampling time, the normalised time-profile and the radiometer equation.
\begin{equation} \label{eq:radiometer}
    F=\sum S/N \frac{\rm{SEFD(\nu,\theta_z)}}{\sqrt{N_p\nu_{\rm{ch}} \,t_{\rm{samp}}}}\times t_{\rm{samp}}
\end{equation}
where $S/N$ is the signal-to-noise ratio per channel, $\rm{SEFD(\nu,\theta_z)}$ is the system equivalent flux density of Arecibo, which is dependent on the observing frequency ($\nu$) and the zenith angle ($\theta_z$),  $N_p$ is the number of polarisations (2 for the Arecibo L-wide receiver), $\nu_{\rm{ch}}$ is the bandwidth of a single channel and $t_{\rm{samp}}$ the sampling time.
In the literature, the gain ($G$) and system temperature ($T_{\rm{sys}}$) of Arecibo have typically been approximated as $10.5$\,K\,Jy$^{-1}$ and $30$\,K, respectively. We have modelled the SEFD ($=T_{\rm{sys}}/G$) as a function of observing frequency and zenith angle, by fitting an exponential function to system performance measurements taken with the L-band wide receiver over a few years\footnote{\url{http://www.naic.edu/\%7Ephil/sysperf/sysperfbymon.html}}. The sensitivity decreases as zenith angle increases and frequency decreases.  This approach improves upon the standard convention of using one average value, as an SEFD $\approx$ 2.9\,Jy (often used in literature) can underestimate the SEFD by $\lesssim25\,$percent on average.

%how we calculate energy
Finally, the isotropic equivalent burst energy\footnote{Even though the emission is expected to be beamed, we default to the isotropic equivalent energy because the beaming fraction is unknown.}, $E$, is calculated as follows

\begin{equation} \label{eq:iso_energy}
    E = \frac{4\pi \textit{F} \Delta\nu \textit{D}_\text{L}^2}{1+z} 
\end{equation}

where $z$ is the redshift, $F$ is the fluence of the burst (calculated only over the time- and frequency-extent of the burst), $\Delta\nu$ is the bandwidth of the burst and $D_\text{L}$ is the luminosity distance to the host galaxy of \frb\ \citep[972\,Mpc;][]{tendulkar_2017_apjl}. 

The fluence of a burst can be calculated either over the entire observing frequency range (often the central observing frequency is used to offset the effect of varying instrumental bandwidths) or only over the extent of the burst. Common practice \citep[e.g.,][]{law_2017_apj} and recent recommendation \citep[e.g.,][]{aggarwal_2021_apjl} has preferred the latter. Importantly, however, when the isotropic equivalent energy is computed, one needs to consider the same frequency range as for the fluence estimation. Inconsistency in this matter may result in certain features of the burst energy distribution being diminished or accentuated.

\subsection{Grouping bursts and comparing groups}

\label{sec:two_groups}

%how we cluster the bursts into 2 groups
In Figure~\ref{fig:e_v_w} the burst widths have been plotted against their isotropic equivalent energy, while also being coloured according to their observed bandwidth within the 1150$-$1730\,MHz range we recorded. Our sample of bursts splits into two distinct groups in this three dimensional parameter space. We refer to these two groups as the low-energy bursts (LEBs) and high-energy bursts (HEBs). We employed a Gaussian Mixture Models (GMM) clustering algorithm using the \texttt{GaussianMixture.fit} method in the \texttt{sklearn.mixture} package\footnote{\url{https://scikit-learn.org/stable/}} \citep[][]{scikit-learn} to determine which bursts belong to which group. GMM clustering algorithms assume that all data points are from a combination of a finite number of Gaussian distributions -- this is not necessarily true for our data, but we only use this clustering method as a quantitative means of classifying bursts as LEBs or HEBs. We disregard bandwidth and only take into account the temporal width and isotropic equivalent energy parameters (both in logarithmic space) for the clustering. We did so, firstly, because the burst divide is most obvious in the width-energy plane. Secondly, the bandwidth is constrained by the maximum observing frequency range. While the bursts in our sample are also restricted to a range of temporal widths (due to boxcar lengths being searched) and energies (due to sensitivity limitations), the bandwidth is the only parameter where an abrupt cut-off is visible in its distribution. We use the isotropic equivalent energy density, which is dependent on the observed burst bandwidth (and, importantly, differs from the intrinsic burst bandwidth). We have refrained from estimating the missing flux/energy outside the observing window since the spectrum of the more complex bursts are non-trivial to model and we believe a one-dimensional Gaussian function to be an oversimplification. If instead, we make use of spectral energy density (erg/Hz), we are able to remove the energy parameter's dependence on bandwidth. Figure~\ref{fig:e_v_w} has been replotted making use of spectral energy density rather than isotropic equivalent energy density in Figure~\ref{fig:SEDvW}. While the separation between LEBs and HEBs diminishes, two groups are still visible.  This metric is also not completely unbiased since, in some instances, bursts extend beyond the observable frequency range of the receiver.  In such cases, we divide by an observed bandwidth that may be lower than the full spectral extent of the burst. We thus adhere to using isotropic equivalent energy, but note that the differences that we see between the HEBs and LEBs are much greater than what one would expect if arising solely from a change in bandwidth (on which the isotropic equivalent energy is dependent). In Figure~\ref{fig:e_v_w} the LEBs and HEBs have been plotted as circles and triangles, respectively. A total of 435 bursts were classified as LEBs and 43 as HEBs. Note that the LEBs typically have lower bandwidths than HEBs, and also larger temporal widths. 

%testing if these groups are truly different
To determine whether the LEBs and HEBs can be drawn from the same parent population, we applied Kolmogorov-Smirnov (KS), Anderson-Darling (AD) and Epps-Singleton (ES) tests using the \texttt{ks\_2samp}, \texttt{anderson\_ksamp} and \texttt{epps\_singleton\_2samp} methods from the \texttt{scipy.stats} package\footnote{\url{https://docs.scipy.org/doc/scipy/reference/stats.html}}. The critical value for the KS test is approximately  $D=c(\alpha)\sqrt{\frac{n_1+n_2}{n_1n_2}}\approx0.26$ where $c(\alpha)=1.63$ is a coefficient corresponding to a significance level ($\alpha$) of 0.01 and $n_1=435$ and $n_2=43$ are the numbers of bursts classified as LEBs and HEBs, respectively. A KS-statistic larger than this value, or, equivalently, a small p-value ($<$0.001) can be used to reject the hypothesis that the samples are drawn from the same parent population. For the ES-test we followed the recommendation of \citet{epps_1986} and set the values where the empirical characteristic function will be evaluated to $t=(0.4,0.8)$. The results of all these statistical tests are tabulated in Table~\ref{tab:ks}. All three tests concur that we can reject the hypothesis that the bandwidth, width and fluence distributions between the two groups can be drawn from the same parent population.

\begin{table}
\caption{The statistical test results from the comparison between LEB and HEB properties}
\label{tab:ks}
\begin{tabular}{llll} \hline \hline
                 & KS statistic & p-value & Bootstrapping p-value\\ \hline
Width            &  0.47            &  3.0 $\times10^{-8}$      & 8.5$\times10^{-8}$ \\
Bandwidth        &  0.56            &  3.8 $\times10^{-12}$      & 1.3$\times10^{-10}$ \\ 
Fluence & 0.63 & 2.0 $\times10^{-15}$ & 1.0$\times10^{-13}$ \\ \hline

    &   AD statistic & p-value & Bootstrapping p-value\\ \hline
    Width & 22.9 & < 0.001 & < 0.001 \\
    Bandwidth & 50.5 & < 0.001 & < 0.001 \\ 
    Fluence & 54.2 & < 0.001  & < 0.001 \\ \hline
    & ES statistic & p-value & Bootstrapping p-value\\ \hline
    Width & 63.3 & 5.9 $\times10^{-13}$ & 3.8 $\times10^{-10}$ \\
    Bandwidth & 86.8 & 6.1 $\times10^{-18}$ & 1.4 $\times10^{-12}$ \\
    Fluence & 39.9 & 4.6 $\times10^{-8}$ & 1.7 $\times10^{-6}$  \\\hline
    
\end{tabular}
\end{table}

%more testing
To ensure that the statistically significant dissimilarity we see between the LEBs and HEBs is not merely the effect of a few outliers, we made use of a quasi- bootstrapping resampling method, whereby we ran 1000 trials, randomly resampling 10 percent of the bursts from both the LEBs and HEBs by drawing from the LEB and HEB distributions, respectively (with replacement). For each trial of each parameter (width, bandwidth and fluence) we calculated the KS, AD and ES statistic and accompanying p-value. The mean p-values of this investigation are also in Table~\ref{tab:ks}, and confirm that the statistical significance of our results is not dominated by a small number of outliers.

\subsection{Burst energetics}
\label{sec:burst_e}
%how we estimate completeness
In order to estimate the completeness limit of our observations, we assume a burst with a S/N of 6 that has a temporal width of 4.00\,ms and a bandwidth of 275\,MHz. These are the median values for width and bandwidth for all the bursts in our sample. Making use of Equations~\ref{eq:radiometer} and \ref{eq:iso_energy}, we can then estimate the fluence or energy threshold below which our burst sample is incomplete. This limit is $\sim0.057 $\,Jy\,ms, or $\sim1.5\times10^{37}\,$erg. Assuming a worst case instead of using the median values quoted above, this limit is increased to $\sim5.5\times10^{37}\,$erg.

%some background on the fitting the energy distribution
The cumulative energy distribution is usually approximated as a power-law $R(>E)\propto E^\gamma$, where $R$ is the burst rate for bursts above some energy $E$ and $\gamma$ is the slope of the power-law. Fitting for a single power-law has proven to likely be an oversimplification when considering data over many orders of magnitude \citep[][]{cruces_2021_mnras}. Moreover, \citet[][]{aggarwal_2021_apjl} have shown that the narrowbandedness of bursts can greatly affect the shape of the observed energy distribution. Because many bursts are cut-off at the top or bottom of the observing window, energy distributions typically do not represent the energy distribution of bursts but rather of the emission within the observing frequency range. One can estimate the missing flux by assuming the spectral shape of the burst (which vary a lot between bursts) or by only selecting the bursts that appear to be fully within the observing window. This, however, also introduces a bias as the brightest broadband bursts will then be excluded. Approximating the energy distribution using slightly more complicated functions, for instance a broken power-law \citep[][]{aggarwal_2021_arxiv} or a combination of a log-normal and Lorentz function \citep[][]{li_2021_natur} have yielded better results. As has been discussed in \citet[][]{gourdji_2019_apjl}, correctly estimating the completeness threshold will also affect the interpretation of the results. 
 
\begin{figure*}
    \centering
    \includegraphics[width=0.85\textwidth]{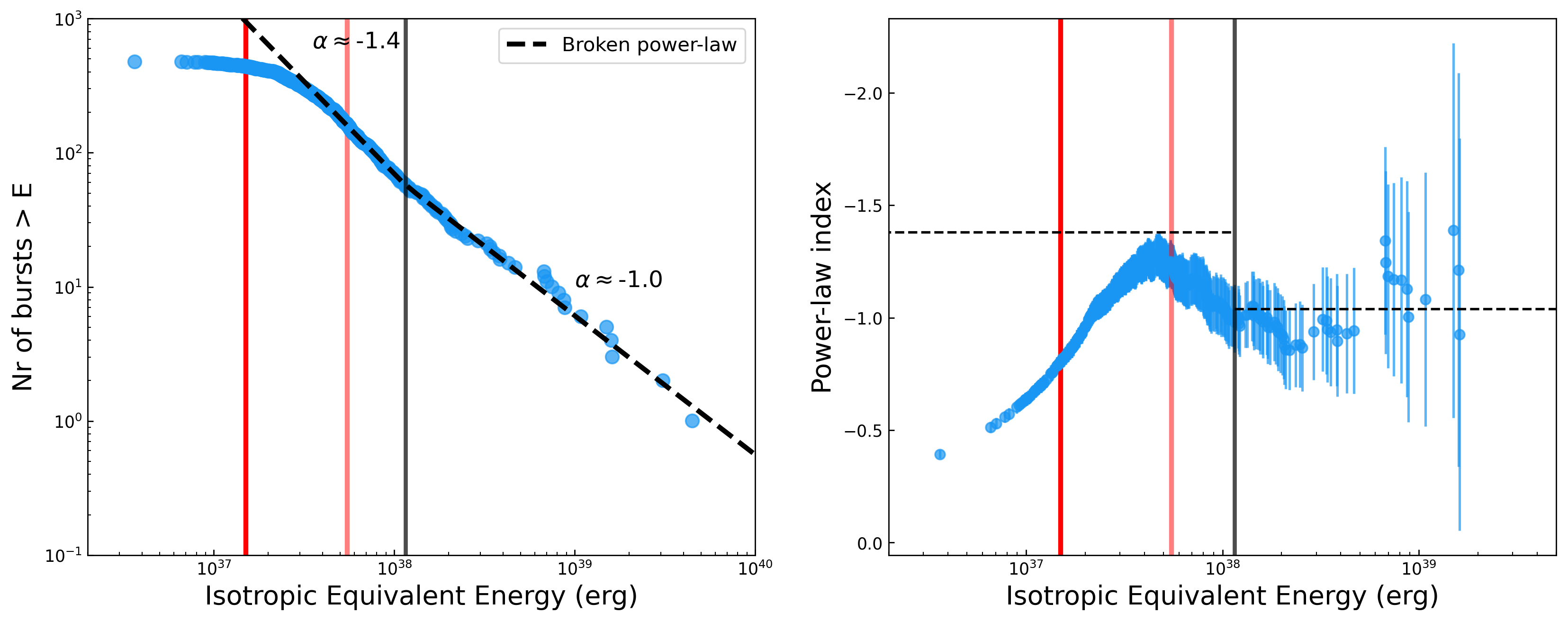}

	\caption{\textbf{Left:} The cumulative energy distribution overplotted with a broken power-law fit that excludes bursts below the conservative sensitivity threshold (5.5$\times10^{37}$\,erg). The vertical red line indicates the sensitivity threshold assuming a burst with a S/N of 6 and the mean burst properties of this sample, while the lighter red line indicates the conservative sensitivity threshold. The black dashed line illustrates the broken power-law fit to the data, with the black vertical line indicating the break of the power-law fit. \textbf{Right:} the power-law index as a function of threshold energy calculated using a maximum-likelihood estimation. The horizontal dashed lines indicate the power-law indices obtained from fitting a broken power-law to the data with a least squares method. Vertical lines are the same as for the left panel.}
    \label{fig:powerlaw}
\end{figure*}

%how we fit the energy distribution
The rate-energy distribution of our sample of bursts can be seen in Figure~\ref{fig:e_v_w}b and Figure~\ref{fig:powerlaw}.
We used \texttt{scipy.optimize.curvefit}\footnote{\url{https://docs.scipy.org/doc/scipy/reference/generated/scipy.optimize.curve_fit.html}} to fit a power-law and a broken power-law to the energy distribution of our sample of bursts and subsets thereof. For the power-law fits, we excluded bursts that are below our calculated energy sensitivity threshold (i.e $E_{\rm{iso}}<1.5\times10^{37}$\,erg), which make up $\sim$8\,percent of the total sample (when fitting the broken power-law we make use of a very conservative sensitivity threshold of $5.5\times10^{37}$\,erg to ensure that we fit the kink in the energy distribution at higher energies and not the turnover due to loss of sensitivity at lower energies). These thresholds are indicated by  red vertical lines in Figure~\ref{fig:powerlaw} and the turn-over at low energies due to our sensitivity loss is also apparent here. The best fit was that of a broken power-law to all the bursts above the conservative sensitivity threshold. This fit yielded a break energy of $E_{\rm{break}}=1.15\pm0.04\times10^{38}$\,erg, with power-law slopes of $-1.38\pm0.01$ where $E_{\rm{iso}} < E_{\rm{break}}$,  and $-1.04\pm0.02$ where $E_{\rm{iso}} > E_{\rm{break}}$ when considering all bursts above the sensitivity threshold. We then repeated the process, only selecting the bursts above our sensitivity threshold that are mostly within the band (the 2$\sigma$ region of the Gaussian fit to the spectrum is within the observing range of $1150-1730$\,MHz) \textit{or} which have a bandwidth larger than 500\,MHz. This is done in an attempt to only account for those bursts for which we measure a significant fraction of the fluence. This yielded $E_{\rm{break}}=1.3\pm0.1\times10^{38}$\,erg, with power-law slopes of $-0.96\pm0.02$ where $E_{\rm{iso}} < E_{\rm{break}}$ and $-0.68\pm0.03$ where $E_{\rm{iso}} > E_{\rm{break}}$. The results of these power-law fits (and power-law fits to the LEBs and HEBs, separately) are presented in Table~\ref{tab:powerlaws}. Furthermore, we estimated the power-law index using the maximum likelihood method for different completeness thresholds, as described by \citet{crawford_1970_apj} and \citet{james_2019_mnras}, but the power-law index does not to converge towards a single value (see the right panel of Figure~\ref{fig:powerlaw}). This suggests that the energy distribution is not well described by a single power-law.

\section{Discussion} \label{sec:discussion}

\subsection{Energy distribution}

%previous studies
The energy distribution of bursts from \frb\ has been studied by various other authors \citep{law_2017_apj,gourdji_2019_apjl,oostrum_2020_aa,cruces_2021_mnras,aggarwal_2021_arxiv,li_2021_natur}. These authors have found different power-law slopes over different energy regimes (summarised in Table~\ref{tab:lband_r1}). Making use of the maximum likelihood method described in \S\ref{sec:burst_e} these authors find power law slopes between $-1.1$ and $-1.8$. Notably \cite{law_2017_apj} find a less steep average power-law index of $\sim-0.7$, irrespective of the burst frequencies (1.4 and 3.0\,GHz bands) and burst rate, but only have a very small sample of bursts. The values we obtain here, for a broken power-law fit, are roughly consistent with those reported by \citet[][]{Jahns_2022_arXiv}, who found power-law indices of $-0.85$ and $-1.47$ above and below a break energy of $1.6\times10^{38}$, respectively, but slightly flatter than what \citet[][]{aggarwal_2021_arxiv} find ($-1.8\pm0.2 > E\approx2\times10^{37}\,$erg) by fitting a broken power-law. The power-law index appears to be steepening from higher to lower energies, before decreasing again as the incompleteness regime is entered. We also observe that the power-law index is potentially changing from day to day, but lack the number of bursts required to concretely quantify this. This behaviour is also seen in Arecibo observations of \frb\ a few years later in 2019 \citep[][]{Jahns_2022_arXiv}. More observations, over longer periods, are required to reveal how the time-dependent nature of the energy distribution --- as seen in \citet[][]{li_2021_natur} --- affects the slope of the energy distribution.

%comparison with other objects
Highly magnetized neutron stars are promising candidates for comparison of burst energy distributions, by reason of the multitude of theories which advocate for them as the progenitors of FRBs. The power-law indices of fits to the energy distribution that we and other authors have found for \frb\ are consistent with voltage scaling models made for magnetars \citep{wadiasingh_2020_apj}. There are a few pulsars, like the Crab pulsar (PSR~B0531+21), that are known to display giant pulses \citep[GPs;][]{heiles_1970_natur,staelin_1970_natur}. GPs are typically very short (sometimes shorter than a ns) and $2-4$ orders of magnitude brighter than normal pulses \citep[e.g][]{hankins_2003_natur, hankins_2007_apj}.  The pulse-energy distribution of the GPs from the Crab pulsar obey a power-law \citep[e.g.,][]{argyle_1972_apjl,lundgren_1995_apj,popov_2007_aa,karuppusamy_2010_aa}, although the power-law index is typically much steeper ($-\alpha\approx2-3$) than for \frb.  It is also worth noting that the V-shape seen in Figures~\ref{fig:e_v_w} and \ref{fig:fluencevswidth} is similar to what is seen for GPs from the Crab pulsar, where the narrower GPs tend to be brighter \citep[][]{karuppusamy_2010_aa,popov_2007_aa}, however, the timescales are much shorter for the Crab pulsar. 

MeerKAT observations of another young pulsar, PSR~J0540$-$6919, yielded hundreds of GPs with a flux distribution that can be fit by a power-law with a very steep slope of $-2.75$.  Interestingly, there are also deviations from the power-law slope that are seen at all observing epochs \citep[][]{geyer_2021_mnras}, similar to the deviations we observe in our power-law fits in this work. Only a small fraction of the GPs ($<3\,$percent) were detected in a fraction of the band, however, implying that these kinks are not the result of narrowbandedness. 
  
 \subsection{High- and low-energy bursts} 
%Energy break + clustering agree
The classification of the bursts into two groups (LEBs and HEBs), as discussed in \S\ref{sec:two_groups}, also appears to be consistent with what we can infer from the shape of the burst energy distribution. The break in the broken power-law fit occurs at approximately the same energy where we find the transition between LEBs and HEBs (see Figure~\ref{fig:powerlaw}).  \citet[][]{li_2021_natur} have reported that the energy distribution of \frb\ can be modelled by a combination of a log-normal and Cauchy distribution. The break in our power-law fit occurs at approximately the same energy as where these two functions in their fit merge ($\sim10^{38}\,$erg).

\begin{figure*}
    \centering
    \includegraphics[width=0.85\textwidth]{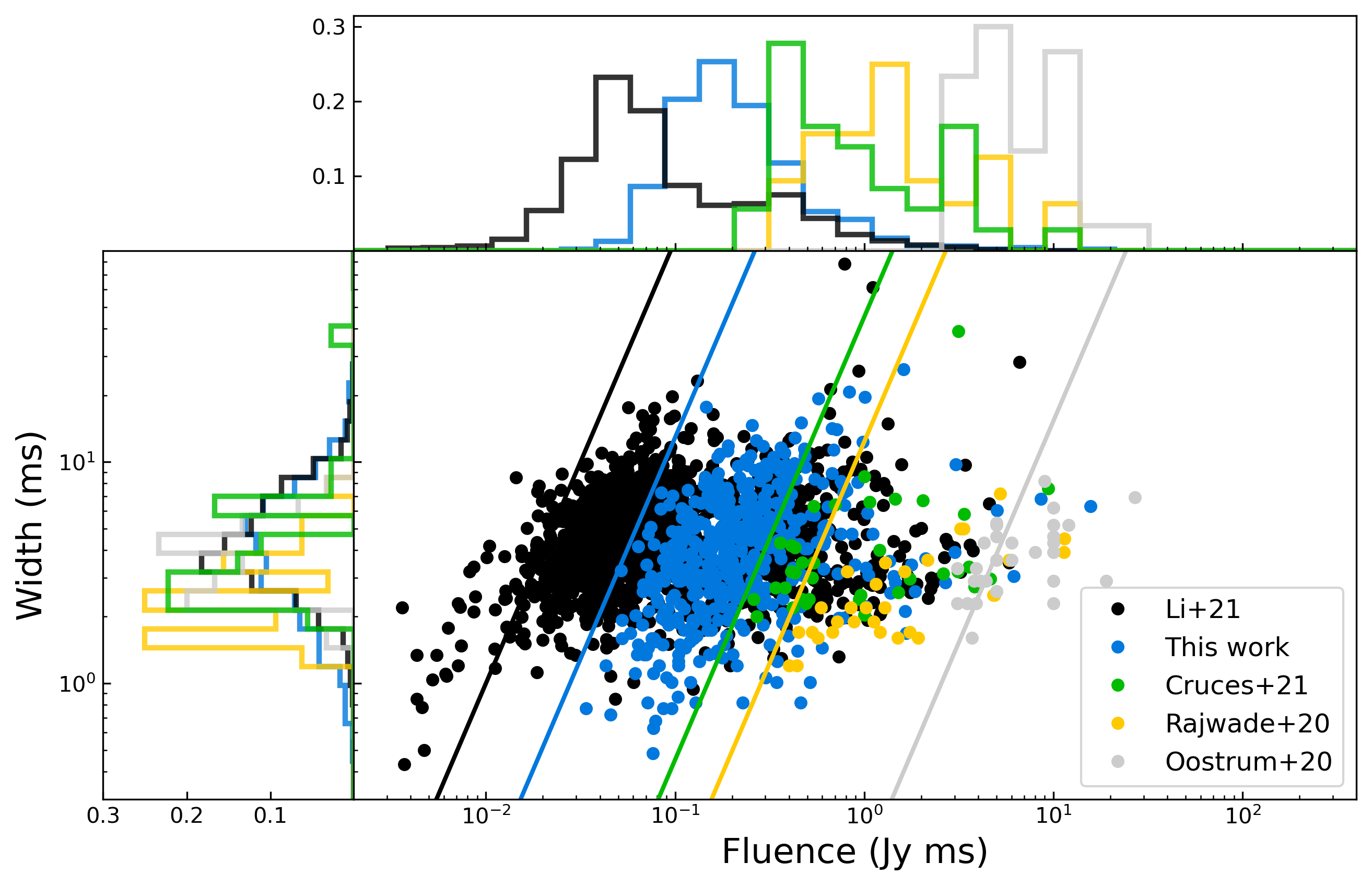}

	\caption{ Width-fluence distribution for our Arecibo-detected bursts in blue, as well as for bursts from FAST \citep[][]{li_2021_natur}, Effelsberg \citep[][]{cruces_2021_mnras}, Lovell \citep[][]{rajwade_2020_mnras} and Apertif \citep[][]{oostrum_2020_aa} in black, green, yellow and grey, respectively. The diagonal lines represent the sensitivity threshold of the respective instruments. Side histograms in the left and top panel show the distribution of the burst width and fluence, respectively. These histograms have been normalised by the number of bursts reported in each of the respective studies. We caution against over-interpretation of the values being compared here, since the definition of `width' and method of calculating the fluence are somewhat inconsistent between samples.}
    \label{fig:fluencevswidth}
\end{figure*}

%these two groups are not separated but bleed into each other
The two potential groups of bursts we have identified are unlikely to inhabit mutually exclusive areas of the parameter space. Instead, the burst rate of high-fluence bursts from both groups is lower than that of the low-fluence bursts: with sufficient observation time and activity from \frb, the LEBs and HEBs will partially overlap in their distributions. This can be seen in Figure~\ref{fig:fluencevswidth} where we plot the fluence against the width of our sample of bursts, as well as bursts from \citet[][]{rajwade_2020_mnras},  \citet[][]{oostrum_2020_aa}, \citet[][]{cruces_2021_mnras} and \citet[][]{li_2021_natur}. The details of these other observations are summarised in Table~\ref{tab:lband_r1}. The forward-slanted V-shape seen in Figure~\ref{fig:e_v_w} can also be seen here, although less distinctly, and the group of LEBs is broader and extends down to lower fluences due to the sensitivity of the FAST observations. We observed for a total of $\sim$59 hours, $\sim$39 hours of which fall within the proposed activity period of \frb\ \citep{rajwade_2020_mnras,cruces_2021_mnras}. In comparison, \citet[][]{cruces_2021_mnras} observed $\sim$128 hours, $\sim$75 hours within the activity window, and \citet[][]{li_2021_natur} observed $\sim60$\,hours, all within the active period. Given that these authors spent more time observing during the active period than we did, this illustrates how, given sufficient observation time, rarer high-fluence LEBs (and HEBs) will be detected and the two groups will appear to merge in the fluence-width parameter space. We caution against over-interpreting the overlap of bursts in this figure, as widths and fluences are not always defined or calculated in a consistent manner.\footnote{\citet[][]{oostrum_2020_aa} and \citet[][]{li_2021_natur} quoted boxcar widths, while \citet[][]{rajwade_2020_mnras}, \citet[][]{cruces_2021_mnras} and this work define the width as the FWHM of a Gaussian fit.} Furthermore, the energy distribution of the bursts from \frb\ has been seen to change over the course of one activity cycle \citep[][]{li_2021_natur}. Moreover, in the case of FRB~20180916B, the burst activity has been shown to drastically change from one cycle to the next \citep{pleunis_2021_apjl}. If the same mechanisms are at play in \frb, this further complicates the comparison of observations during different activity windows. Finally, we note that the classification of the lowest-energy bursts with the smallest temporal widths is particularly ambiguous, and can be influenced by the energy-metric used, since the LEBs and HEBs overlap in this part of the parameter space.

\begin{table*}

\caption{Observation properties of \frb\ bursts at L-band}
\begin{tabular}{llllll} \hline \hline
Telescope  & SEFD & Observing Frequency & Observing & Energy range & Power-law\\  
 &  (Jy) & Range (MHz) & Time (hrs) & (erg) & index\\ \hline
Apertif$^a$    & 700  & 1130 -- 1760  &            $\sim$130            & $\sim 7 \times 10^{38} - 6 \times 10^{39} $ & $-1.7\pm0.6$ \\
Arecibo (This work)   & 3.5  & 1150 -- 1730              & $\sim$57                 & $\sim 4 \times 10^{36} - 8 \times 10^{39} $ & See Table~\ref{tab:powerlaws} \\
Arecibo$^b$    & 2.9  & 1150 -- 1730              & $\sim$3                 & $\sim 7 \times 10^{36} - 2 \times 10^{38} $ & $-1.8\pm0.3$\\
Effelsberg$^c$ & 17   & 1210 -- 1510              & $\sim$128             & $\sim 1 \times 10^{38} - 1 \times 10^{39}$ & $-1.1\pm0.1$\\
FAST$^d$     &  1.25    & 1050 -- 1450              & $\sim$60           & $\sim 4 \times 10^{36} - 8  \times 10^{39} $ & $-0.61\pm0.04$ \\ 
&&&& $\sim 3 \times 10^{38} - 8  \times 10^{39} $ & $-1.37\pm0.18$\\
Lovell$^e$     &   38   & 1400 -- 1800              & $\sim$198           & -  & - \\ \hline
\end{tabular} \\
$^a$\citet[][]{oostrum_2020_aa}
$^b$\citet[][]{gourdji_2019_apjl}
$^c$\citet[][]{cruces_2021_mnras} $^d$\citet[][]{li_2021_natur}, in this case the power-law fit is to the isotropic equivalent energy distribution rather than the cumulative-rate energy distribution. $^e$\citet[][]{rajwade_2020_mnras}
\label{tab:lband_r1}
\end{table*}

\begin{figure}
    \centering
    \includegraphics[width=0.45\textwidth]{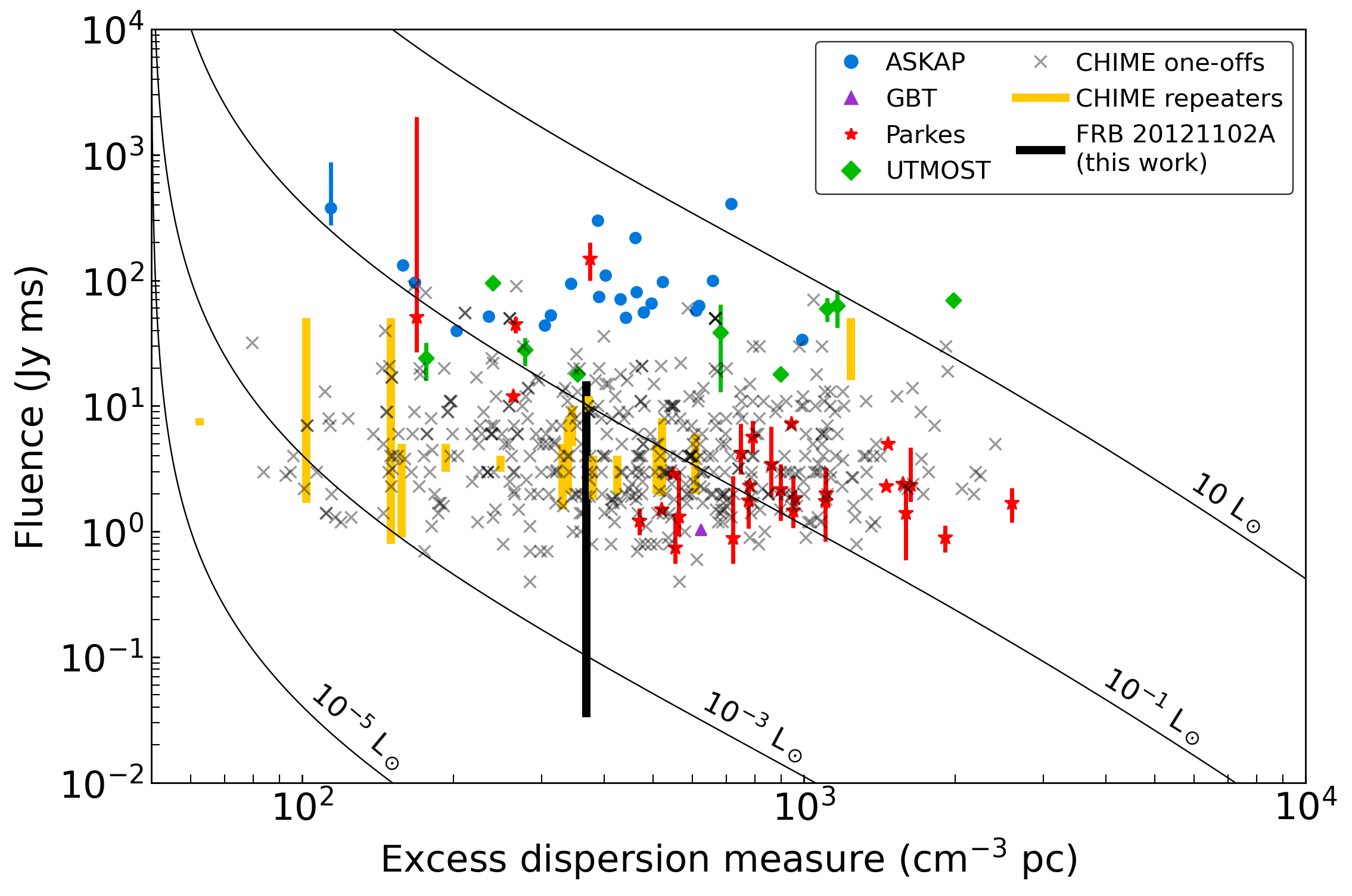}

	\caption{Fluence plotted against excess dispersion measure (corrected for the inferred Galactic foreground contribution) for the sample of bursts from \frb\ detected in this paper (where the black vertical line indicates the range of fluences), as well as various other one-off and repeating FRBs. The different colours represent FRBs detected by different telescopes. Yellow bars indicate the range of fluences detected for CHIME/FRB repeaters. Lines of constant luminosity, assuming a DM contribution from the host galaxy DM$_{\rm{host}}$ = 50\,cm$^{-3}\,$pc, are overplotted. Figure adapted from \citet[][]{shannon_2018_natur,petroff_2019_aarv}. }
    \label{fig:fluencevsdmesc}
\end{figure}

%HEBs look like 1offs and LEBs like repeaters
In a comparison of burst fluences and luminosities, \citet[][]{shannon_2018_natur} found that FRBs detected by Parkes were on average of lower fluence, but higher DM than those detected by ASKAP. These different distributions arise from different telescope sensitivities and instantaneous field-of-view.  Nonetheless, the Parkes and ASKAP FRBs have comparable luminosities (using DM as a proxy for distance). Additionally, the energies of bursts originally detected from \frb\ were relatively low, compared to ASKAP and Parkes one-offs. This may lead one to speculate that the one-offs are in fact the most energetic bursts detected from repeaters. These FRBs are all plotted in an adapted version of the fluence-dispersion plot in Figure~\ref{fig:fluencevsdmesc} \citep[][]{shannon_2018_natur,petroff_2019_aarv}.  The properties of bursts from the two groups we identify in \S\ref{sec:two_groups} in fact correspond to some of the differences observed between one-off and repeating FRBs from the CHIME/FRB catalogue \citep[][]{chime_2021_arxiv_arxiv210604352}, possibly further supporting this notion: HEBs and CHIME/FRB one-offs are typically broader band and shorter duration than LEBs and CHIME/FRB repeater bursts.

%counter-arguments
The analogy is, however, not perfect. \citet[][]{chime_2021_arxiv_arxiv210604352} have found no evidence to suggest that the fluence distributions of the bursts from the one-offs and repeaters are different (both CHIME/FRB catalogue one-offs and repeaters are also plotted in Figure~\ref{fig:fluencevsdmesc}). Neverthelesss, these fluence distributions might not fully be representative of the underlying populations, as it can also be seen that the majority of current instruments lack the sensitivity to probe burst fluences below approximately 1\,Jy\,ms. Should the known repeating FRBs be LEB equivalents and the observed one-offs be HEB equivalents, one would not necessarily expect them to have different fluence distributions. Rather, the burst fluence distributions of known repeaters would extend into the territory of one-offs. The brightest bursts from \frb\ are comparable in luminosity to the one-offs detected by ASKAP and Parkes, and the faintest bursts extend beyond the sensitivity threshold of all but FAST (and, previously, that of Arecibo).

Another potential caveat arises when considering a sample of 41 low-energy bursts from \frb\ (almost all of which are also presented here) presented by \citet[][]{gourdji_2019_apjl}, who  showed that the narrowbandedness of fainter bursts might be explained by receiver sensitivity that is insufficient to detect fainter sub-components. This argument suggests that the LEBs are merely the `tips of the iceberg', but on the other hand HEBs possess smaller temporal widths in general than LEBs, contradicting this notion. 

The strongest potential counter-argument comes from the accumulating evidence in the literature  that supports polarimetric differences between repeaters and one-offs. If all FRBs are repeaters and known repeaters are akin to LEBs while known one-offs are the HEBs, the polarimetric properties of LEBs and HEBs are expected to be different. We do not perform any polarimetric studies as the frequency resolution of our data is insufficient to resolve the high RM of \frb\ at our observing frequency, but others have found no polarisation from any \frb\ bursts at $\sim$1.4\,GHz \citep[][]{li_2021_natur}. Polarisation is, however, detected at $\sim4.5$\,GHz \citep[][]{michilli_2018_natur}, so the lack of polarisation at $\sim$1.4\,GHz could potentially be due to the extreme local environment of the source imparting depolarisation at lower frequencies \citep[][]{Plavin_2022_arXiv}. Multi-path propagation in a scintillation screen between the FRB source and observer can cause linear depolarisation, as well as impart significant circular polarisation \citep[][]{beniamini_2022_mnras}.

\subsection{Burst diversity and the average burst spectrum}

\begin{figure}
    \centering
    \includegraphics[width=0.45\textwidth]{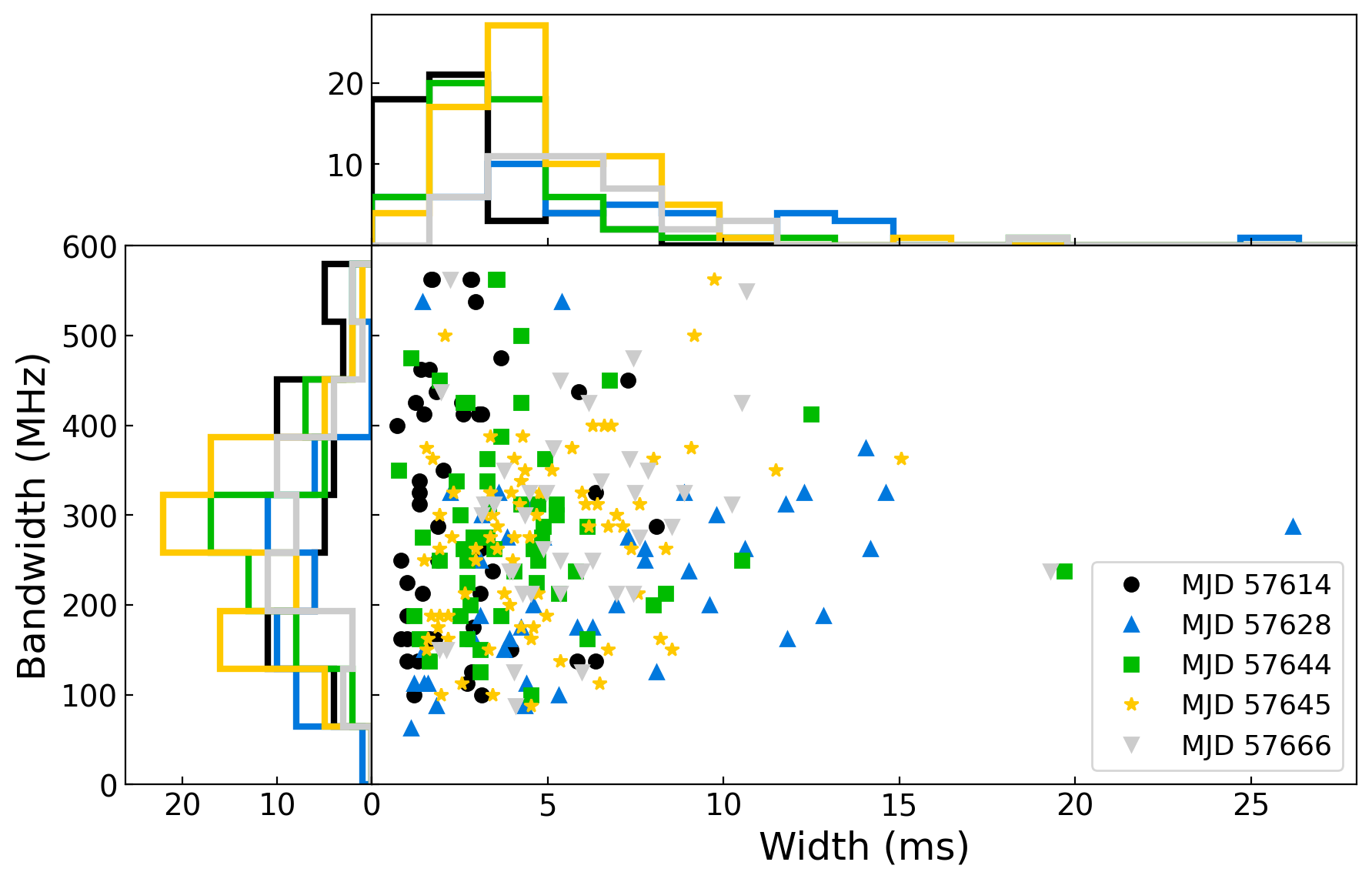}

	\caption{Bandwidth-width distribution of bursts from our sample on days where more than 40 bursts were detected. Different days are indicated by different colours (see the MJD in the legend). Side histograms in the left and top panels show the observed bandwidth and width distributions, respectively.}
    \label{fig:daytodayvariation}
\end{figure}

Already at the time of \frb's discovery as the first known repeating FRB, \citet[][]{spitler_2016_natur} remarked on the high variability between burst spectra. Many others have also noted the diversity in both the time and frequency properties of the bursts. That being said, the burst morphology of some of the more complex bursts is remarkably similar: e.g., the precursor bursts observed by \citet[][]{caleb_2020_mnras} or the similarity between bursts AO-02 and GB-01 in \citet[][]{hessels_2019_apjl}. Whether this is a physically meaningful observation or mere coincidence remains to be seen.
This diversity in the time and frequency properties of the bursts is showcased in Figure~\ref{fig:dynamic_spectra}. The narrowest and widest bursts we detected are 0.5 (B205) and 26.0\,ms (B107), respectively. The bursts also extend over different frequency ranges, with some being very narrowband (as narrow as $\sim$\,65\,MHz), while others extend over the whole observing bandwidth of 580\,MHz. We found no evidence for periodic variation in any of the burst properties (see Appendix Figures~\ref{fig:multiplot1}-\ref{fig:multiplot4}). We do, however, observe that the bursts from the same epoch are, on average, more similar to each other than to bursts from other epochs. The irregular burst rate makes this effect especially difficult to quantify. In Figure~\ref{fig:daytodayvariation} the observed bandwidth of bursts are plotted against temporal widths for days where more than 40 bursts were detected. One can see, e.g., that a larger fraction of the bursts from MJD~57614 have larger bandwidths and that the bursts are of shorter duration compared to MJD~57628. For MJD~57614 and MJD~57628 the median bandwidths and widths are 292\,MHz and 1.95\,ms and 223\,MHz and 5.08\,ms, respectively. This might suggest an emission region that is evolving over the course of days, but stable over a period of $\sim$hours.  However, the statistical significance of these apparent similarities of burst properties within an observing epoch requires confirmation.

\begin{figure}
    \centering
    \includegraphics[width=0.45\textwidth]{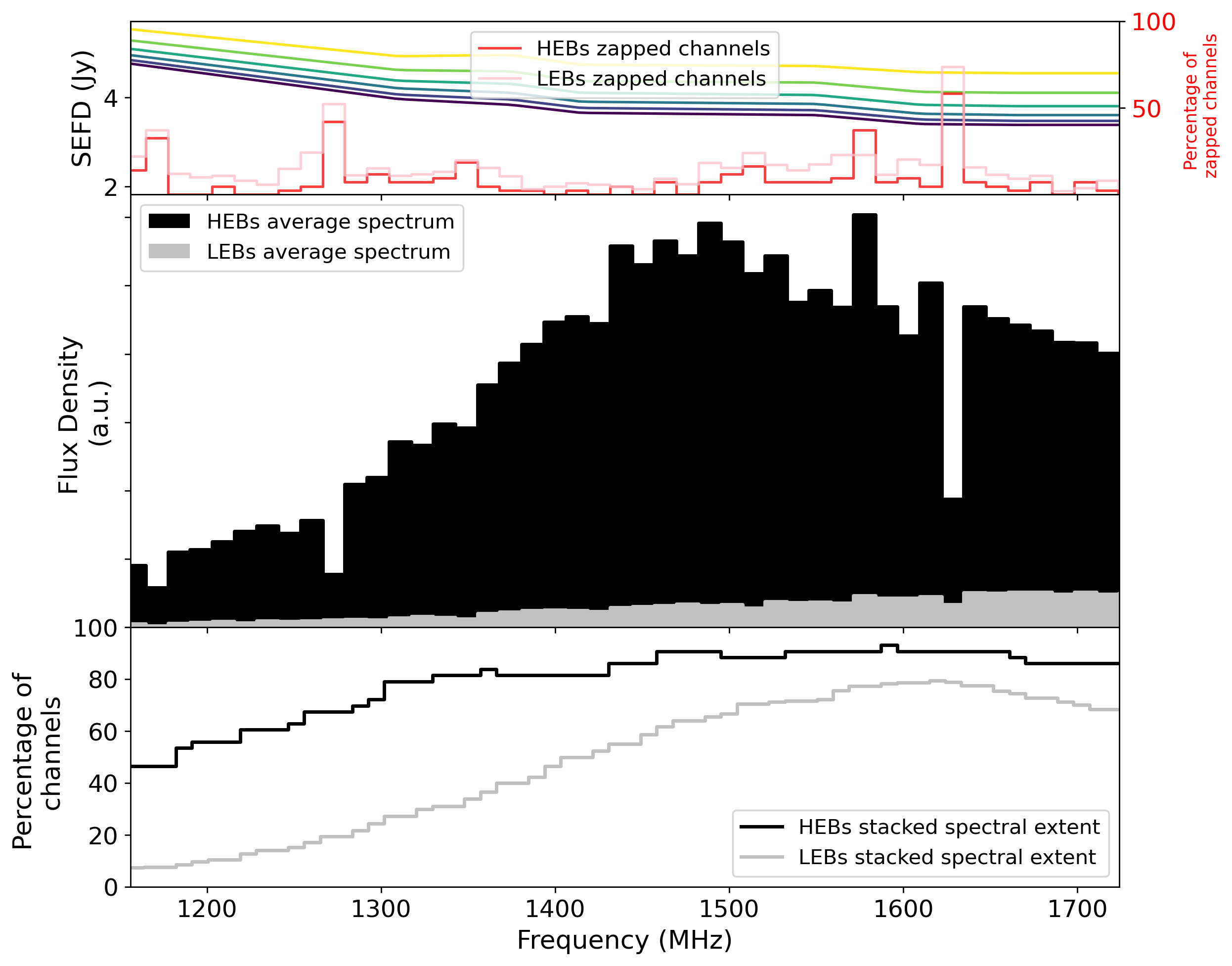}

	\caption{\textbf{Top panel:} The observational limitations: the red and pink histogram shows how many times each frequency channel was flagged due to the presence of RFI for the HEBs and LEBs, respectively. This is given as a percentage of the total number of bursts on the right hand y-axis.  The SEFDs for zenith angles from 14 to 19 degrees (the range covered by \frb\ as it is transiting at Arecibo), in steps of 1 degree, are shown on the left hand y-axis. \textbf{Middle panel:} The spectra obtained from averaging all the spectra for the HEBs (in black) and LEBs (in grey). \textbf{Bottom panel:} The percentage of bursts for which emission was present in a given channel for the HEBs in black and LEBs in grey. }
    \label{fig:spectrum}
\end{figure}

Figure~\ref{fig:spectrum} shows two different methods for determining the average spectrum of the bursts in our sample. In the middle panel the mean spectrum of the 435 LEBs and 43 HEBs detected from \frb\ have been plotted as black and grey histograms, respectively. These S/N-weighted average spectra are obtained by simply summing the spectra together of all the bursts in a group and then dividing by the number of bursts in that group. This plot shows that, on average, we observed factors of approximately 5 more burst emission at higher observing frequencies than at lower observing frequencies, for both the LEBs and HEBs. All spectra have been corrected for bandpass variations and the effect of decreasing sensitivity at lower frequencies has already been accounted for by modelling the SEFD of Arecibo as a function of frequency and zenith angle (see \S\ref{sec:results}).

The top panel in Figure~\ref{fig:spectrum} shows the SEFD as a function of frequency for various zenith angles ranging from 14 degrees in dark blue to 19 degrees in yellow (the range covered by \frb\ as it is transiting at Arecibo). Throughout the observation some channels were flagged more than once due the presence of RFI (manifesting as sharp dips in the average spectra in the middle panel). This decrease in flux density that we observe at lower frequencies is thus not an effect of RFI excision.

The vast majority of bursts occur at the top of our observing band. We define the spectral extent of a burst as the 2-sigma region of a one-dimensional Gaussian fit to the time averaged spectrum of each burst. In the bottom panel of Figure~\ref{fig:spectrum} we stacked the spectral extents of all the bursts to produce an average `spectrum' not weighted by signal intensity. This histogram thus indicates, given a burst detection, at which frequencies the burst emission is most likely visible. Here one can see that fewer bursts are detected at lower frequencies. For the LEBs, the shape is distinctly Gaussian-like and peaks at around 1600\,MHz while dropping off quickly below 1400\,MHz.  Similarly, \cite{gourdji_2019_apjl} noted a ``dearth'' of bursts below 1350\,MHz. The difference in the bandwidths between the LEBs and HEBs noted before is also apparent here. The HEBs are typically broader band. We found that the peak of the summed spectral extent remains more or less constant across our observations (see Appendix Figures~\ref{fig:multiplot1}--\ref{fig:multiplot4}). We suspect that most, if not all, of the bursts detected at our highest observing frequencies extend beyond the observing band. The average stacked spectral extents presented here are thus not a representation of the average burst, but rather the frequency-dependent burst activity within our observing window.

\cite{gajjar_2018_apj} plotted the peak flux density of a number of bursts in the literature as a function of frequency over a much larger frequency range of $\sim 1-8\,$GHz, and found a relatively flat spectral index with no evidence of a spectral turnover. Both \cite{gajjar_2018_apj}, as well as \cite{zhang_2018_apj}, do however find that the bursts they detected between 4 and 8\,GHz seem to have preferred frequencies. This effect is, at least partly, due to Galactic scintillation. Our observations, made at lower frequencies where the Galactic scintillation bandwidth is much narrower \citep[][]{hessels_2019_apjl}, show that the bursts from \frb\ occur at preferential frequencies (that likely evolve with time, see \citet[][]{Jahns_2022_arXiv}).

\subsection{Activity window}

\begin{figure*}
    \centering
    \includegraphics[width=0.9\textwidth]{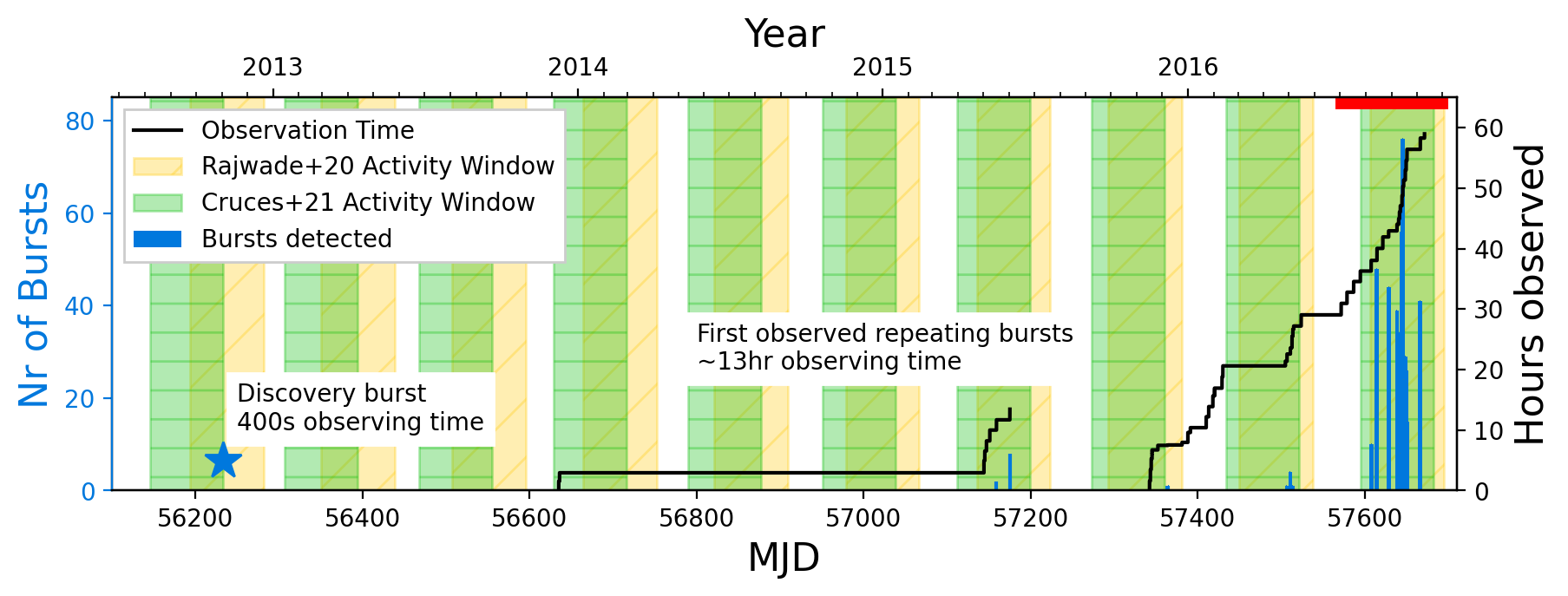}
    \includegraphics[width=0.9\textwidth]{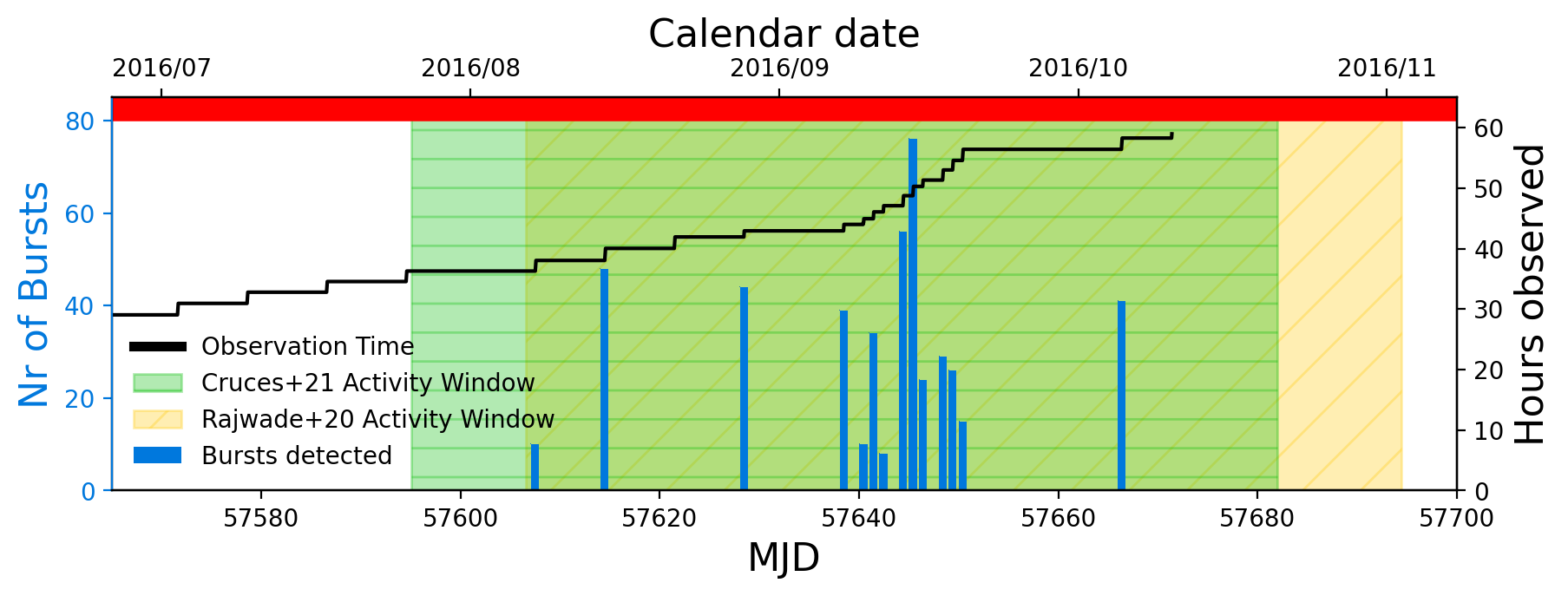}

	\caption{Timeline of Arecibo monitoring of \frb, showing all the \frb\ bursts detected with Arecibo at $\sim1.4\,$GHz up until the end of 2016. MJD is indicated on the bottom axes and calendar date on the top axes. The blue star indicates the date of the first burst ever to be detected from \frb\  \citep{spitler_2014_apj}. Blue histogram bars indicate the number of bursts detected in different observations (left axes). Note the first ten repeat bursts in mid-2015 that were presented in \citet[][]{spitler_2016_natur} and the first burst that was detected with PUPPI at the end of 2015 and  presented in \citet[][]{scholz_2016_apj}.  The cumulative time spent observing is shown in black (right axes), and is reset to zero on MJD~57342, which is when the observations started that we report on in this paper. The tentative activity windows proposed by \citet[][]{rajwade_2020_mnras} and \citet[][]{cruces_2021_mnras} are shown as yellow and green shaded areas, respectively. The bottom panel is a zoom in of the top panel, at the dates indicated by the thick horizontal red bar. Bursts from MJD~57644 and MJD~57645 were also presented in \citet[][]{gourdji_2019_apjl} and \citet[][]{aggarwal_2021_arxiv}. }
    \label{fig:monitoring}
\end{figure*}

In Figure~\ref{fig:monitoring} we illustrate the long-term monitoring of \frb, since its discovery, at $\sim$1.4\,GHz with the Arecibo 305-m telescope.  We also highlight the activity cycles proposed by \citet[][]{rajwade_2020_mnras} and \citet[][]{cruces_2021_mnras}. \citet{rajwade_2020_mnras} proposed a period of 157\,days with a 56\,percent duty cycle, whereas the period from \citet{cruces_2021_mnras} is slightly different: $161\pm5$ days with a duty cycle of 54\,percent. 
After its discovery in 2012 \citep[][]{spitler_2014_apj}, \frb\ was monitored again in 2013, and later in 2015 which resulted in the discovery of the first repeat bursts detected from an FRB \citep[][]{spitler_2016_natur}. Long-term Arecibo monitoring commenced around December 2015, but temporarily ceased early in 2016. Observations again resumed towards the end of this first activity window of 2016, during which a few bursts were detected. The vast majority of the bursts presented in this paper come from the burst storm detected during the second active period in 2016. During this period and after confirmation of  burst detections, we increased the frequency of our observations to approximately daily cadence starting from around $\sim$MJD~57640. The increased cadence, and consequently increased number of burst found, is evident in Figure~\ref{fig:monitoring}. Now, aware of the activity window  of \frb, the lack of detections early in the project can easily be explained and supports the case for an activity window of $\sim$161 days. We do, however, note that a subset of the bursts presented here --- and also in \citet[][]{gourdji_2019_apjl}, \citet[][]{aggarwal_2021_arxiv} and \citet[][]{hessels_2019_apjl} --- were used together with other data to determine this activity window. Continued monitoring of \frb\ in the coming years is necessary to refine this activity window and to determine its long-term stability or variability.

Any progenitor model for \frb\ naturally needs to be able to explain this fairly long period activity window and large duty cycle. \citet[][]{xu_2021_apj} have shown that a supernova fallback disc with the most reasonable values for a disc wind can slow down isolated neutron stars to spin periods of hundreds of hours. A period similar to that of \frb\ can be obtained by invoking an extremely large disc wind, but this challenges what may be physically possible around a magnetar. \citet[][]{wada_2021_apj} have shown that the large activity window of periodic FRBs can be explained by incorporating eccentricity in the binary comb model of \citet[][]{ioka_2020_apjl}. Furthermore, specifically for \frb, and assuming that its persistent radio counterpart is in fact powered by a disc wind, a supermassive or intermediate-mass black hole companion is preferred over a massive stellar companion.

\subsection{Wait times }

\begin{figure}
    \centering
    \includegraphics[width=0.45\textwidth]{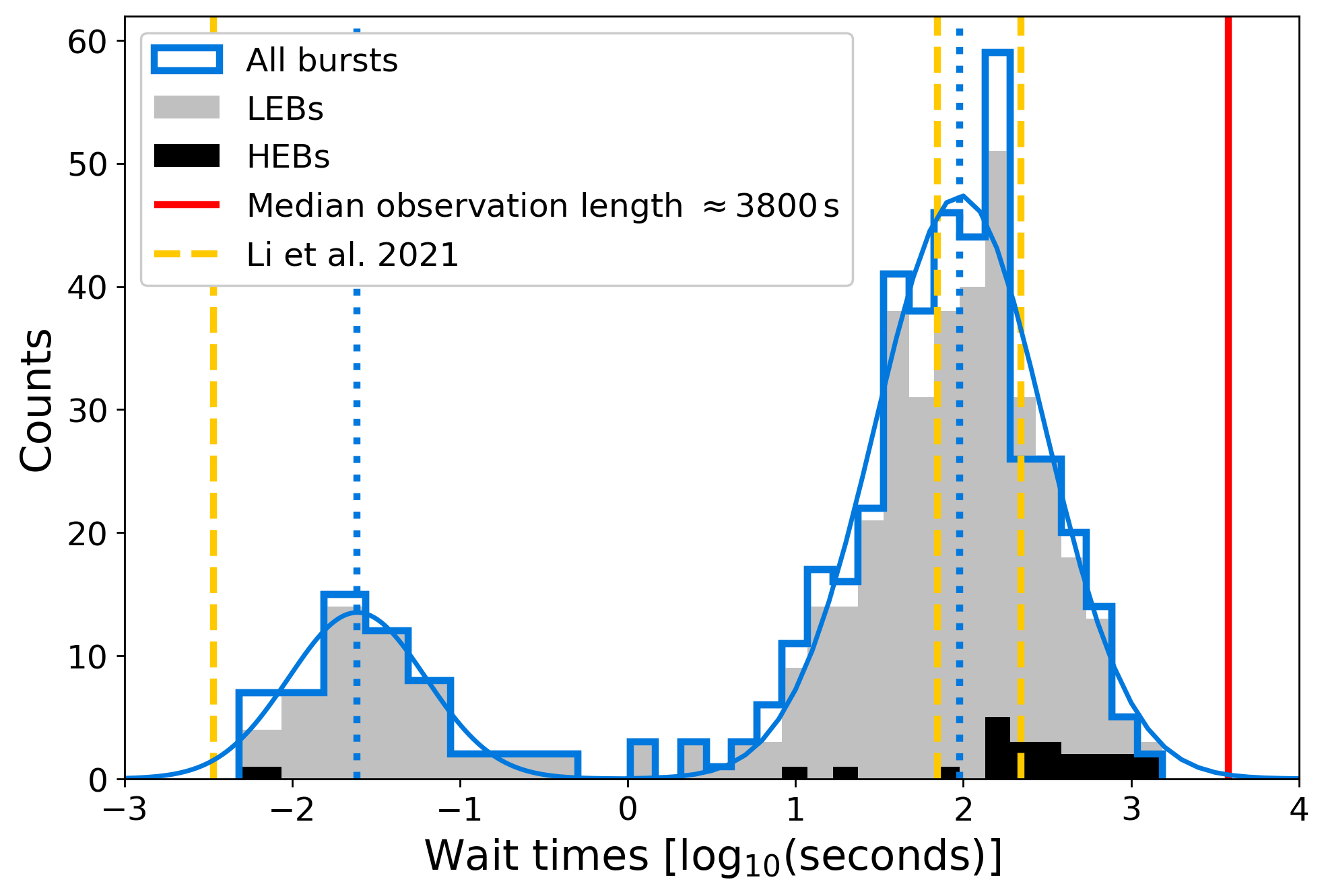}

	\caption{The wait time distribution for all detected bursts is shown by the blue outline. The separate wait time distributions for the LEBs and HEBs are indicated by solid grey and black histograms, respectively. Log-normal functions are fit to the full sample distribution above and below 1 second, and peak at $\sim$24\,ms and $\sim$95\,s, as indicated by the dotted blue lines. Our sensitivity to longer wait times is degraded by the length of our observations. The median length of our observations is indicated by the red vertical line. For comparison, the peaks of the waiting times identified by \citet[][]{li_2021_natur} are indicated by yellow dashed vertical lines at $3.4\pm1.0$\,ms, $70\pm12$\,s and $220\pm100\,$s. The latter being the peak for the higher energy bursts ($E>3\times10^{38}\,$erg) they observe.}
    \label{fig:waittimes}
\end{figure}

Despite its long-term periodic activity, \frb\ has shown no short-timescale periodicity in the arrival times of the bursts like that seen from pulsars or FRB~20191221A \citep{chime_2021_arxiv_arxiv210708463}\footnote{This FRB is a single 3\,s event with sub-second periodicity between \textit{sub-bursts}.}, despite various searches using a range of techniques \citep[][]{cruces_2021_mnras,aggarwal_2021_arxiv,zhang_2018_apj}.  Other authors have found that the wait times between bursts form a bi-modal distribution with peaks at milli- and decaseconds \citep[][]{katz_2019_mnras,gourdji_2019_apjl,li_2021_natur,aggarwal_2021_arxiv}. The wait-time distribution between bursts is dependent on the definition of a burst, and on sub-second scales will look especially different depending on whether the wait times between sub-bursts ($\sim\,$ms) are included or not. We have calculated the wait times for our entire sample of bursts (and not between sub-bursts) and also see two humps in the distribution. We fit the two humps with a log-normal function, by first excluding wait times $<1$\,s and then $>1$\,s. The log-normal functions peak at $\sim$24\,ms and $\sim$95\,s, which is comparable to what has been found by other authors \citep[][]{zhang_2018_apj,aggarwal_2021_arxiv}. The wait time distribution and log-normal fits are presented in Figure~\ref{fig:waittimes}. The durations of our observations are tabulated in Table~\ref{tab:observations}. The median observation duration is $\sim3800$\,s and implies that we are not sensitive to wait times longer than a few kiloseconds. Additionally, we also calculated the wait times separately for the HEBs and LEBs. The HEBs show no sign of clustering in time (see also Appendix Figures~\ref{fig:multiplot1}--\ref{fig:multiplot4}) and they typically have longer wait times, consistent with their lower rate.   

 The peak at longer timescales is consistent with what is expected from a Poissonian distribution. \citet[][]{aggarwal_2021_arxiv} have shown how this decaseconds peak shifts to shorter timescales exponentially as more bursts are detected in a constant-length observation. Our sub-second peak, in combination with the absence of wait times around one second, potentially describes the typical event duration. While we do not intentionally measure the wait times between sub-bursts, at least some fraction of the sub-second wait times are likely the wait times between non-subsequent sub-bursts, where the faintest sub-bursts are not being detected. A `sub-burst' peak is very evident in \citet[][]{li_2021_natur} at $\sim3$\,ms, but not here. In their wait time distribution a second sub-second peak can be seen (but is not explicitly mentioned) corresponding to the one we observe. The sub-second peak we detect is also consistent with those found by \citet[][]{gourdji_2019_apjl}, \citet[][]{aggarwal_2021_arxiv} and \citet[][]{li_2021_natur}, and supports the claims made by the latter authors that challenge the feasibility of certain coherent emission models that require orderly magnetic field lines. 

This bi-modality in the wait time distribution is also seen in magnetars \citep[e.g.,][]{Huppenkothen_2015_ApJ}. In the specific case of SGR~J1550$-$5418, the wait times are between the spikes (akin to sub-bursts) seen in the magnetar X-ray bursts and the peak at lower wait times is around 50\,ms. While \citet[][]{Huppenkothen_2015_ApJ} note that these observed wait times are consistent with both repeated crust failure as well as magnetospheric reconnection models, \citet[][]{wadiasingh_2020_apjl} argue that the shorter timescale observed in the wait time distribution of magnetars is likely associated with magnetar oscillations. It is also worth mentioning that SGR~1935+2154, the only Galactic magnetar known to show an FRB-like burst, has been shown to exhibit (X-ray) burst rates and a fluence power-law index consistent with that observed from \frb, although the wait times between bursts are $\sim2\,$s \citep[][]{Younes_2020_ApJL}. We encourage better quantification of the wait time distribution for other repeating FRBs, as well as searches for periodicities in the wait time distributions.

\subsection{Periodicity search }

A Lomb-Scargle\footnote{\url{https://docs.scipy.org/doc/scipy/reference/generated/scipy.signal.lombscargle.html}} periodicity search was performed on the barycentric corrected burst times of arrival for every observation that contained at least 24 bursts. We tested $2 \times 10^{6}$ trial periods which were logarithmically spaced between $20$\,ms and $200$\,seconds. We find no significant periods that are consistent between multiple observations. Additionally, a fast folding algorithm (FFA) was applied to the observations that contain at least 40 bursts using \texttt{RIPTIDE} \citep{morello_2020_mnras}. As with the Lomb-Scargle search, periods between $20$\,ms and $200$\,seconds were searched. We again find no significant periods that are consistent between multiple observations.  If the central engine of the FRBs is rapidly spinning, then the lack of detected periodicity in the burst arrival times suggests that the emitting region is not stable in location.

\section{Summary} \label{sec:conclusions}

In this paper we have reported the detection of 478 bursts (of which 333 are previously unreported) from \frb\ that were obtained through monitoring with the Arecibo telescope at L-band, between December 2015 and October 2016. Our observations total $\sim59$\,hours, and reflect the earliest large observing campaign of this iconic source. Our analysis is summarised as follows:
\begin{enumerate}
    \item The bursts in our sample are not all identical, and divide into two groups in the width-bandwidth-energy parameter space.  We refer to these two burst types as LEBs (low-energy bursts) and HEBs  (high-energy bursts).
    \item Statistical comparison tests concur that the temporal width, observed bandwidth and fluence distributions for the HEBs and LEBs cannot be drawn from the same parent population.
    \item The HEBs are typically narrower and more broadband than the LEBs, echoing the differences between one-off and repeating FRBs.
    \item We fit the energy distribution of our sample with a broken power-law that has a break at $E_{\rm{break}}=1.15\pm0.04\times10^{38}$\,erg, with power-law slopes of $-1.38\pm0.01$ where $E_{\rm{iso}} < E_{\rm{break}}$, and $-1.04\pm0.02$ where  $E_{\rm{iso}} > E_{\rm{break}}$.
    \item  We observed that bursts from the same epoch tend to be more similar to each other than to bursts from other epochs, but note that this observation requires confirmation.
    \item  The majority of the bursts in our sample occur towards the top of our observing band ($\sim1730$\,MHz) and like previous authors we note a dearth of bursts detected below $\sim1350\,$MHz.
    \item Our observations (including those with non-detections) support the case for a $\sim$160\,day activity period with a $\sim$55\,percent duty cycle.
    \item We find two log-normal peaks in the wait time distribution at $\sim24$\,ms and $\sim95$\,s, consistent with previous measurements and indicative of a characteristic event duration and Poissonian rate, respectively.
    \item Despite using various methods, we find no short-term periodicity in the burst arrival times when searching between $20$\,ms and $200$\,seconds.
\end{enumerate}

\section*{Acknowledgements}

\textit{Facility:} Arecibo \\
\textit{Software:} Astropy, FETCH, DSPSR, PSRCHIVE, PRESTO, psrfits\_utils, RIPTIDE

We would like to thank the referee for thoughtful comments that helped to improve the quality of the manuscript. We thank E. Petroff for providing the code required to make Figure 5.
Research by the AstroFlash group at University of Amsterdam, ASTRON and
JIVE is supported in part by an NWO Vici grant (PI Hessels; VI.C.192.045). LGS is a Lise Meitner Max Planck Research Group Leader and acknowledges funding from the Max Planck Society.
%%%%%%%%%%%%%%%%%%%%%%%%%%%%%%%%%%%%%%%%%%%%%%%%%%
\section*{Data Availability}

The Arecibo data used in this work is available upon reasonable request.



\bibliographystyle{mnras}
\bibliography{arecibor1} % if your bibtex file is called example.bib

%\begin{thebibliography}{99}
%\end{thebibliography}

%%%%%%%%%%%%%%%%%%%%%%%%%%%%%%%%%%%%%%%%%%%%%%%%%%

%%%%%%%%%%%%%%%%% APPENDICES %%%%%%%%%%%%%%%%%%%%%

\appendix
\section{Fit functions}
We define a broken power law as 
\begin{equation}
    N(>E)=\begin{cases}
        k(E/E_\mathrm{break})^{\alpha_1}, & E<E_\mathrm{break}\\
        k(E/E_\mathrm{break})^{\alpha_2}, & E\geq E_\mathrm{break}
    \end{cases}
\end{equation}
where $k$ is a scaling factor and $\alpha_1$ and $\alpha_2$ are the power-law indices below and above the break energy, $E_{\rm{break}}$, respectively.

\section{Extra material}

\onecolumn
\small
\begin{longtable}{lllllllll}
\caption{Burst properties (full table available online)}\label{tab:allbursts} \\ \hline
Burst ID & TOA$^{a}$ & Width (ms) & Bandwidth (MHz) & $f_{\rm{high}}$ (MHz) & $f_{\rm{low}}$ (MHz) & Fluence$^{b}$ (Jy ms) & Energy$^{b}$ (erg) & Group \\
\hline
\endfirsthead
\caption{Burst properties (continued) }\\
\hline
Burst ID & TOA & Width (ms) & Bandwidth (MHz) & $f_{\rm{high}}$ (MHz) & $f_{\rm{low}}$ (MHz) & Fluence (Jy ms) & Energy (erg) & Group \\
\hline
\endhead
\hline
\endfoot
\hline \hline
\endlastfoot
1  &  57364.204636306  & $ 0.68 \pm 0.05 $ &  $> 546$  &  $> 1730$  &  1184  &  0.08  & $ 4.0 \times 10 ^{  37 }$  &  HEB \\
2  &  57506.802409024  & $ 4.7 \pm 0.3 $ &  $> 381$  &  $> 1730$  &  1349  &  0.38  & $ 1.3 \times 10 ^{  38 }$  &  LEB \\
3  &  57510.807219384  & $ 2.27 \pm 0.09 $ &  $> 546$  &  $> 1730$  &  1184  &  0.29  & $ 1.4 \times 10 ^{  38 }$  &  HEB \\
4  &  57510.819153558  & $ 2.2 \pm 0.3 $ &  $> 305$  &  $> 1730$  &  1425  &  0.11  & $ 2.9 \times 10 ^{  37 }$  &  LEB \\
5  &  57510.826409121  & $ 3.4 \pm 0.2 $ &  $> 419$  &  $> 1730$  &  1311  &  0.27  & $ 1.0 \times 10 ^{  38 }$  &  LEB \\
6  &  57510.838080436  & $ 2.0 \pm 0.2 $ &  $> 267$  &  $> 1730$  &  1463  &  0.19  & $ 4.3 \times 10 ^{  37 }$  &  LEB \\
7  &  57513.786194552  & $ 6.9 \pm 0.4 $ &  $> 508$  &  $> 1730$  &  1222  &  0.38  & $ 1.7 \times 10 ^{  38 }$  &  LEB \\
8  &  57607.531624205  & $ 2.7 \pm 0.2 $ &  $> 331$  &  $> 1730$  &  1399  &  0.30  & $ 7.3 \times 10 ^{  37 }$  &  LEB \\
9  &  57607.532581875  & $ 1.1 \pm 0.2 $ &  $> 217$  &  $> 1730$  &  1513  &  0.09  & $ 1.5 \times 10 ^{  37 }$  &  LEB \\
10  &  57607.532581931  & $ 0.8 \pm 0.1 $ &  $> 217$  &  $> 1730$  &  1513  &  0.10  & $ 1.6 \times 10 ^{  37 }$  &  LEB \\
... & ...               & ...             & ...       &  ...       & ...    &  ...   & ...                        & ... \\

\end{longtable}
$^{a}$ These times are dynamical times (TDB), corrected to the Solar System Barycenter to infinite frequency assuming a dispersion measure of 560.5 pc\,cm$^{-3}$ and dispersion constant of 1/(2.41$\times$10$^{-4}$ ) MHz$^{2}$\,pc$^{-1}$\,cm$^{3}$\,s.
 $^{b}$ We estimate a conservative $25$\,percent uncertainty on these measurements due to uncertainty in the SEFD of Arecibo.

\begin{table}
\caption{Power-law fit parameters} \label{tab:powerlaws}
\begin{tabular}{llllll} \hline \hline
Single power-law        & $\alpha_1$       &                            &                                            &                & $R^2$    \\  \hline
All bursts              & $-0.87\pm0.01$   &                            &                                            &                & 0.961 \\
LEBs                    & $-0.98\pm0.01$   &                            &                                            &                & 0.943 \\
HEBs                    & $-0.46\pm0.02$   &                            &                                            &                & 0.956 \\
In-band bursts only     & $-0.72\pm0.01$   &                            &                                            &                & 0.978 \\ \hline
Broken power-law        & $\alpha_1$       & $E_{\rm{break}}$              & $\alpha_2$                                 &                & $R^2$    \\ \hline
All bursts              & {$-1.38\pm0.01$} & $1.15\pm0.04\times10^{38}$ & $-1.04\pm0.02$                           &                & 0.999 \\
LEBs                    & $-1.94\pm0.02$ & $-$ & $-$  &                          &                 $-$ \\
HEBs                    & $-0.39\pm0.02$  & $3.2\pm0.2\times10^{38}$   & $-0.89\pm0.04$  &                                          & 0.993 \\
In-band bursts only     & $-0.96\pm0.02$   & $1.3\pm0.1\times10^{38}$ & $-0.68\pm0.03$ &                                          & 0.995 \\ \hline \hline
\end{tabular}
\end{table}

\begin{figure}
    \centering
    \includegraphics[width=0.4\textwidth]{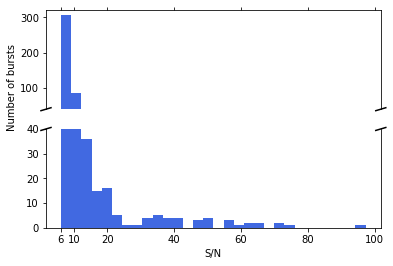}

	\caption{The S/N of all the bursts detected by PRESTO's \texttt{single\_pulse\_search.py}. This is the S/N of the boxcar with the highest S/N at the DM that maximises S/N. }
    \label{fig:sn}
\end{figure}

\begin{figure}
    \centering
    \includegraphics[width=0.85\textwidth]{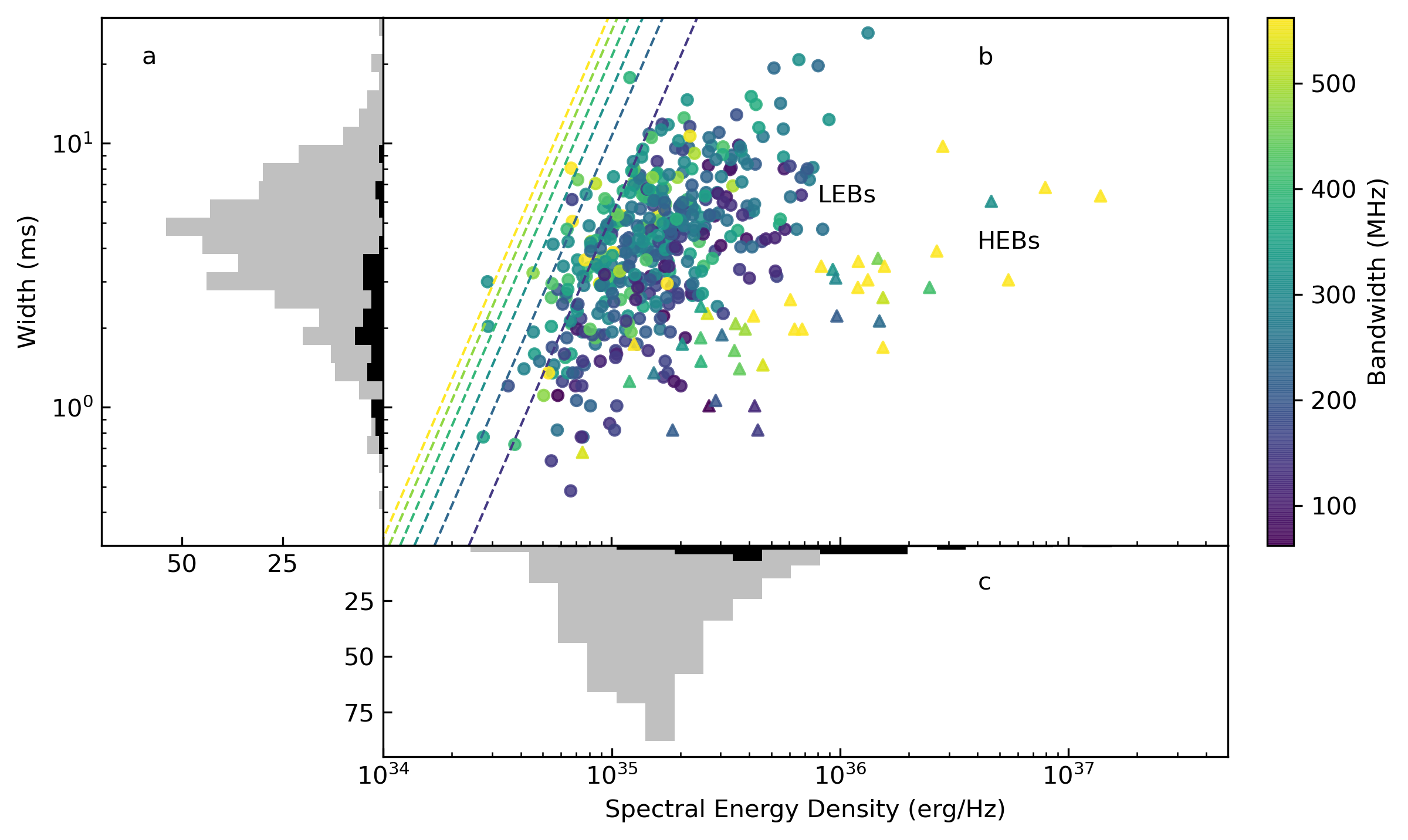}

	\caption{Figure 2 replotted with isotropic equivalent energy (erg) replaced by spectral energy density (erg/Hz). The classification of the lowest energy bursts with the smallest temporal widths becomes ambiguous, because the LEBs and HEBs overlap.  }
    \label{fig:SEDvW}
\end{figure}

\begin{figure*}
	\includegraphics[width=1\textwidth]{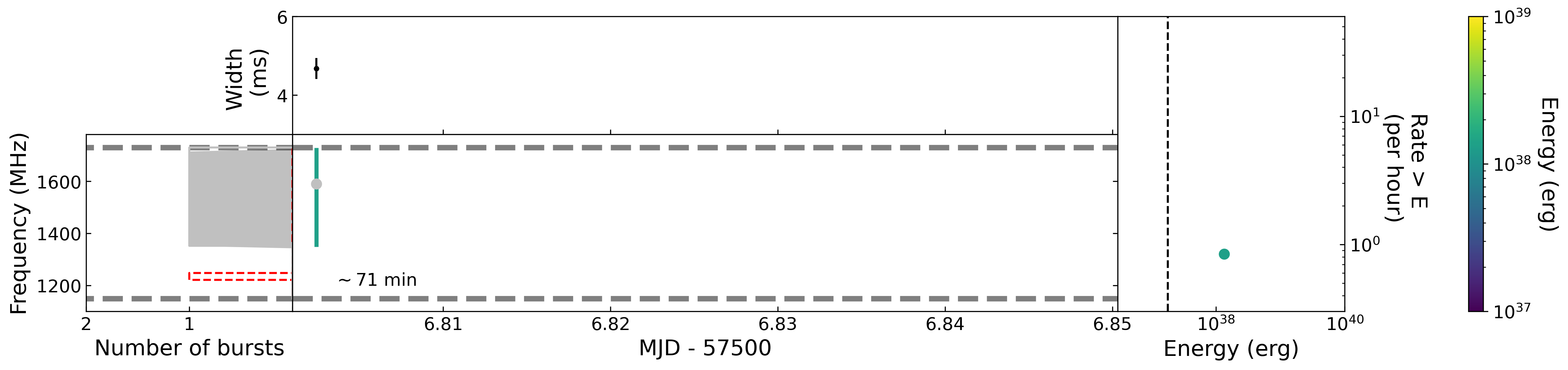}
    \includegraphics[width=1\textwidth]{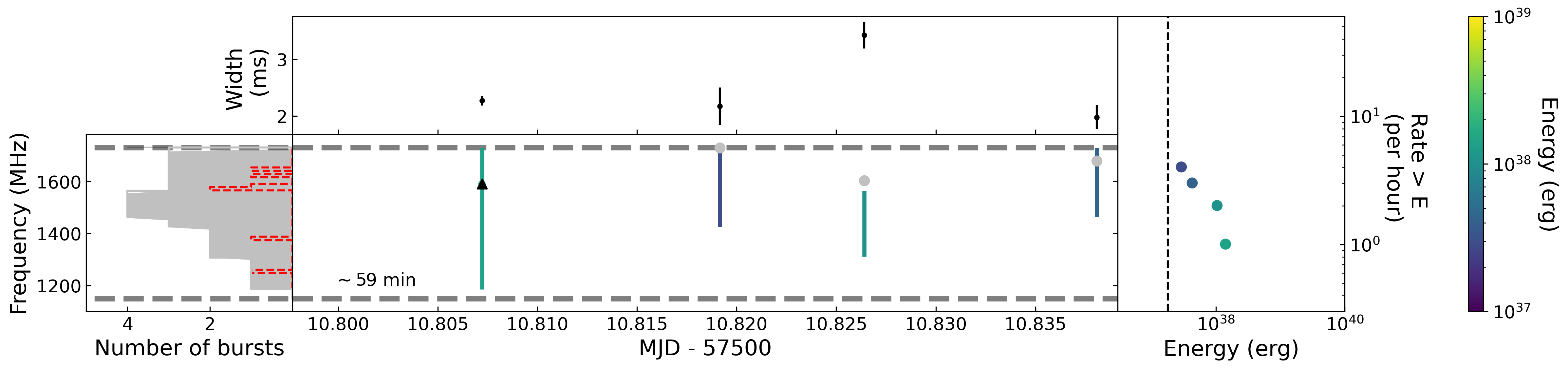}
    \includegraphics[width=1\textwidth]{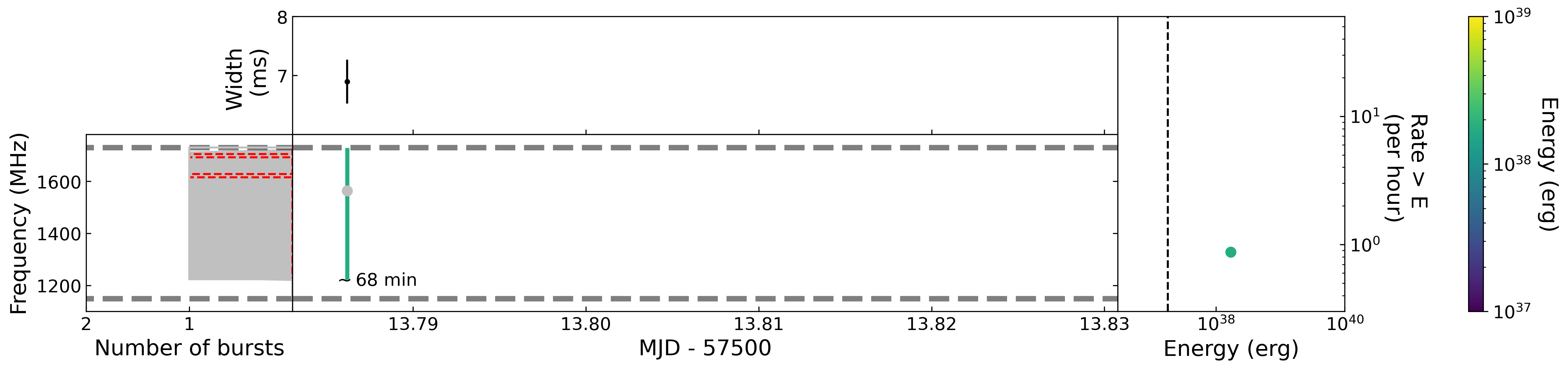}
    \includegraphics[width=1\textwidth]{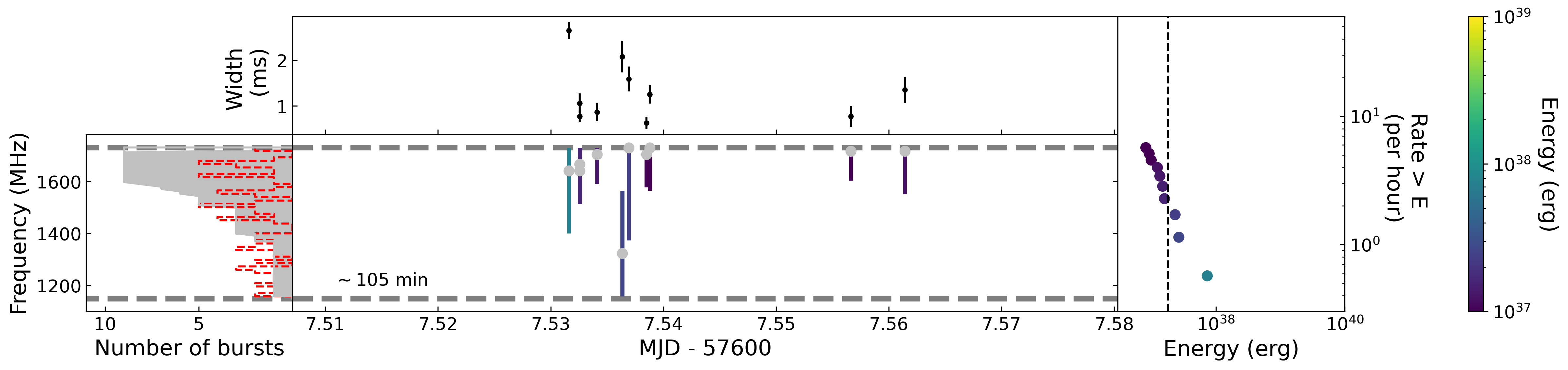}

	\caption{In Figures~\ref{fig:multiplot1} -- \ref{fig:multiplot4} the burst properties have been illustrated for all days on which bursts were detected. In the top panel the temporal width of the bursts is plotted as a function of the MJD. The range of the MJD x-axes are representative of the duration of each observation, and the scale changes between plots of different days. In the center bottom panel a vertical line, coloured according to the calculated isotropic equivalent energy of the burst, shows the spectral extent of each burst as a function of MJD. A grey circle or black triangle, indicates whether a burst is classified as a LEB or HEB, respectively, and is plotted on top of the line at the frequency corresponding to the peak of the Gaussian fit to the spectrum. All the vertical lines are stacked upon each other and displayed in the grey histogram in the left panel, illustrating the frequencies where a burst is most likely to be found during the observation. The overplotted red, dotted histogram indicates for how many bursts specific channels were flagged. Horizontal dashed grey lines at the top and bottom of the bottom panels represent the upper and lower edges of the observing band (1150 and 1730\,MHz). Finally, in the right hand panel a burst energy cumulative distribution of the observation is shown, which has been colour-coded according to energies. These colours also correspond to the data points in the bottom center panel. Here we show MJDs 57506, 57510, 57513 and 57607. }
    \label{fig:multiplot1}
\end{figure*}

\begin{figure*}
    \includegraphics[width=1\textwidth]{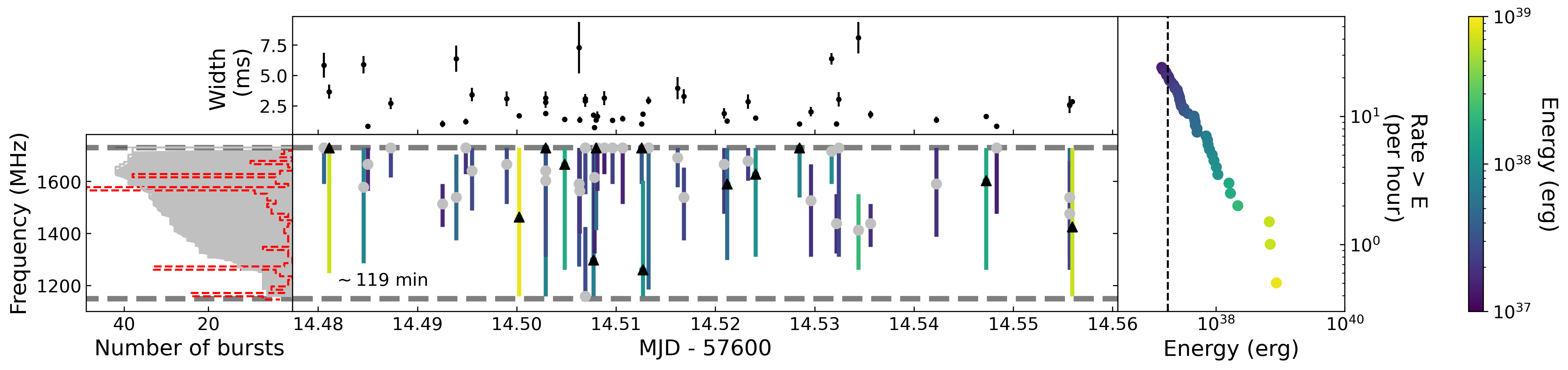}
    \includegraphics[width=1\textwidth]{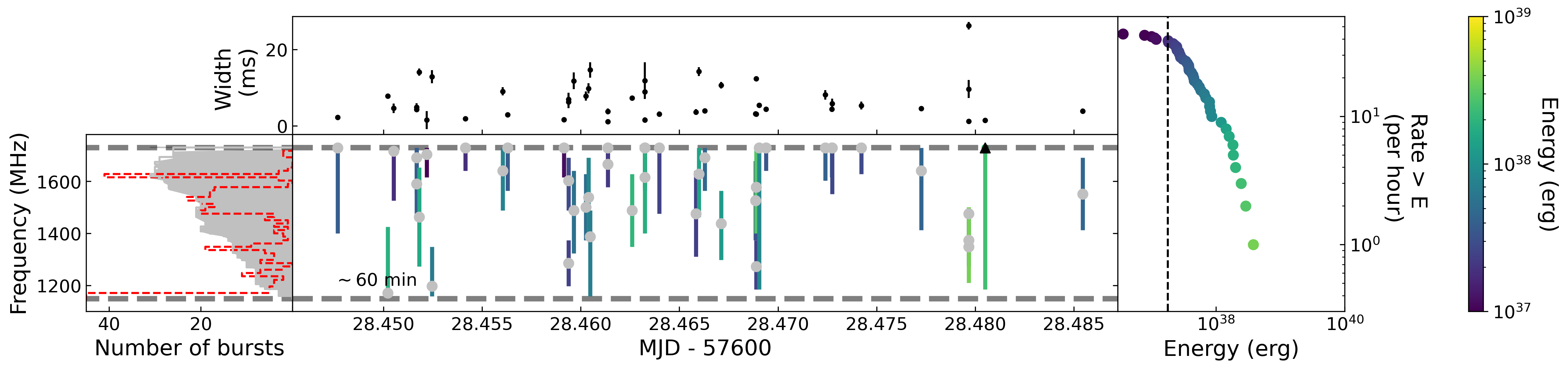}
    \includegraphics[width=1\textwidth]{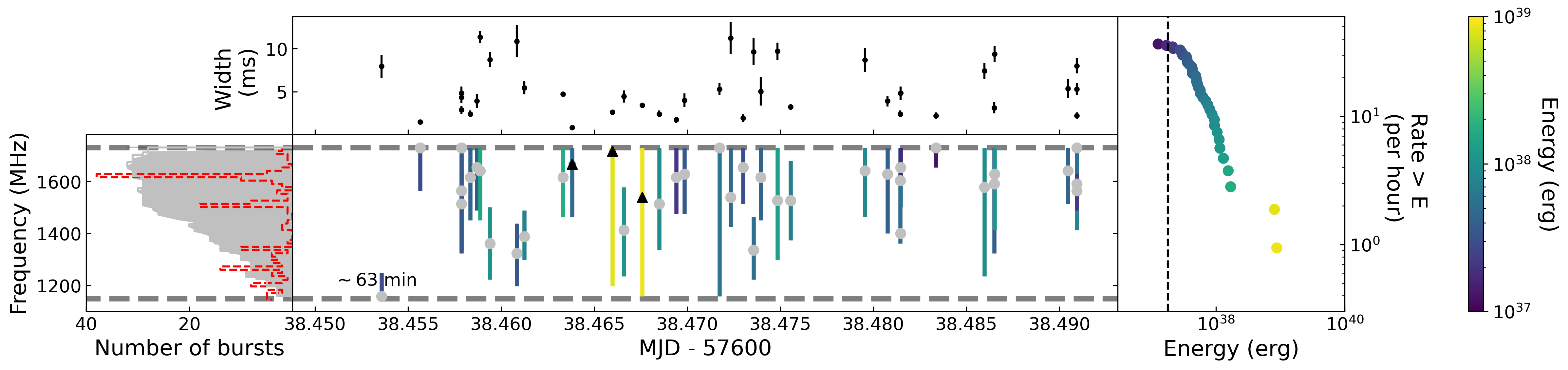}
    \includegraphics[width=1\textwidth]{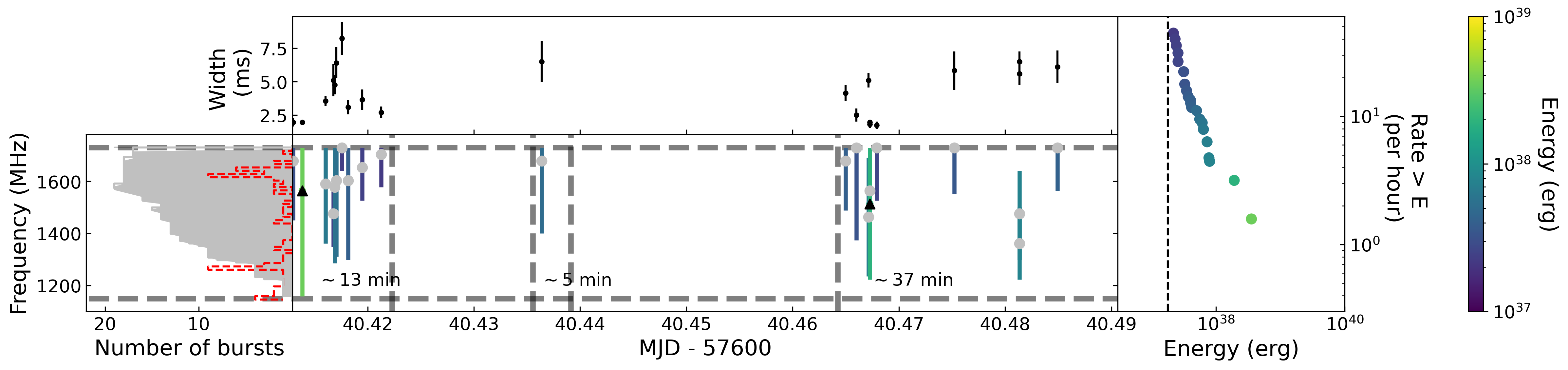}
    \includegraphics[width=1\textwidth]{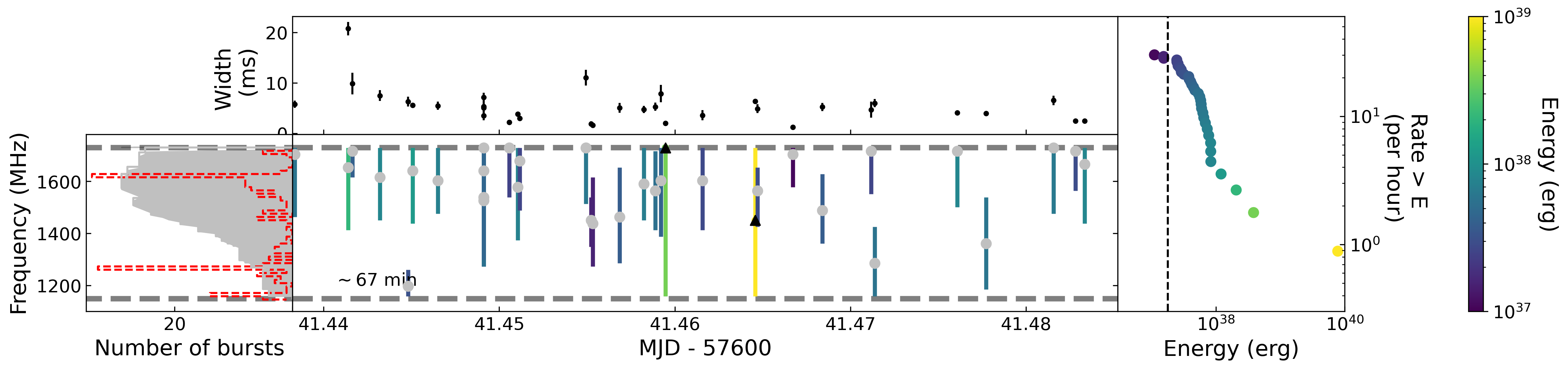}
    
	\caption{Same as  \ref{fig:multiplot1}, but for MJDs 57614, 57628, 57638, 57640 and  57641.}    \label{fig:multiplot2}
\end{figure*}

\begin{figure*}
    \includegraphics[width=1\textwidth]{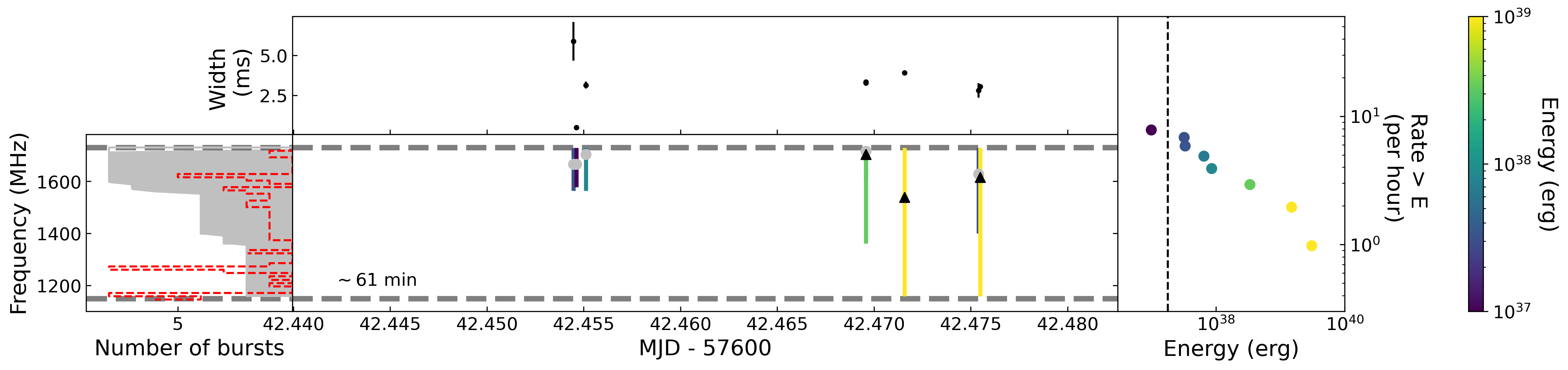}
    \includegraphics[width=1\textwidth]{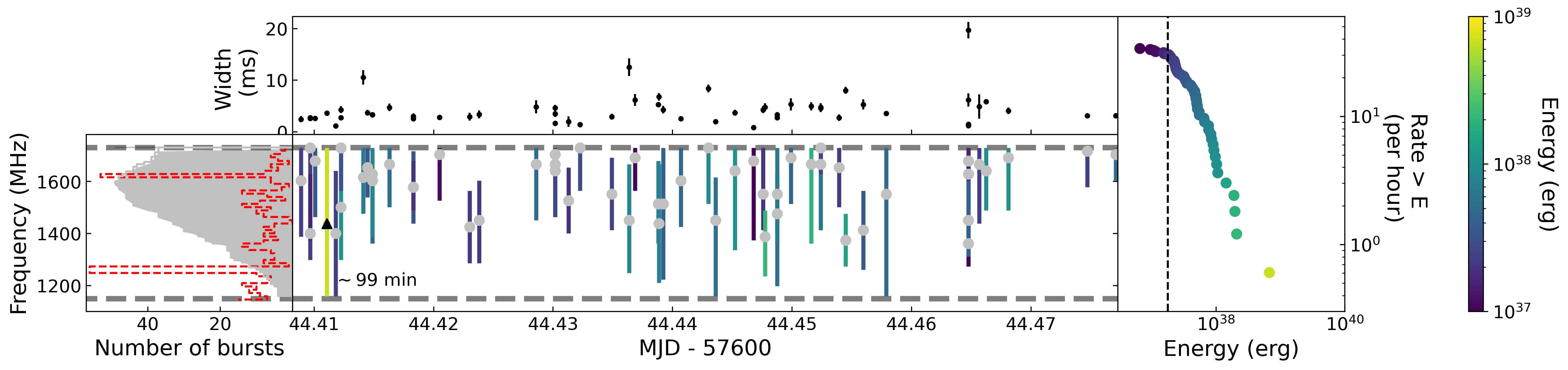}
    \includegraphics[width=1\textwidth]{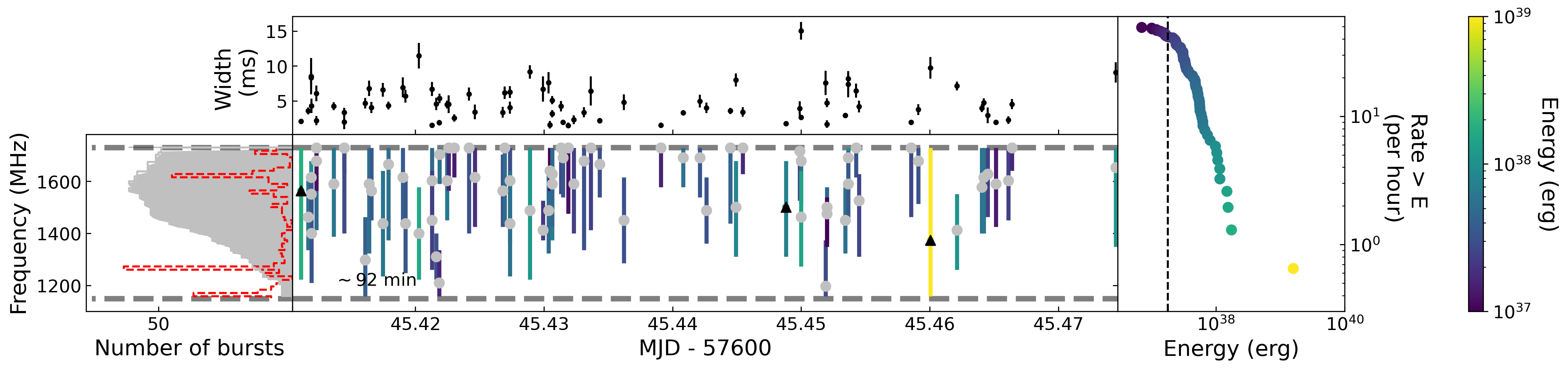}
    \includegraphics[width=1\textwidth]{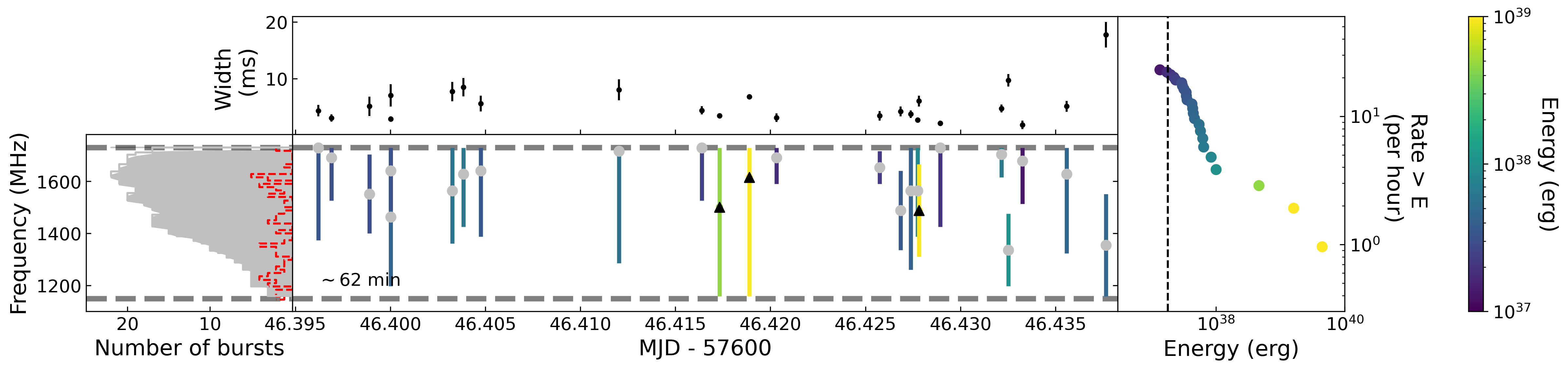}
    \includegraphics[width=1\textwidth]{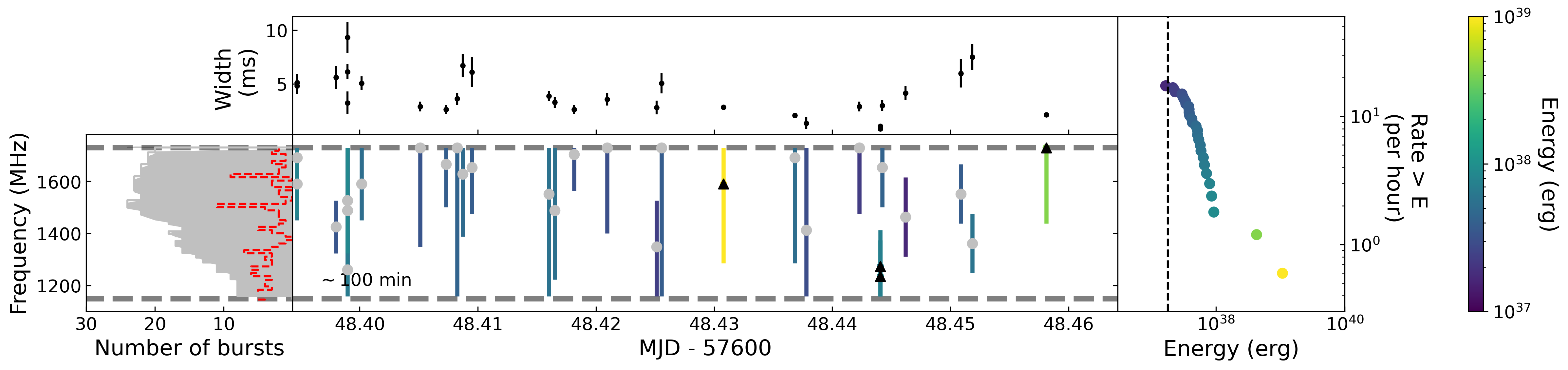}

	\caption{Same as \ref{fig:multiplot1}, but for MJDs 57642, 57644, 57645, 57646 and 57648.}
    \label{fig:multiplot3}
\end{figure*}

\begin{figure*}
    \includegraphics[width=1\textwidth]{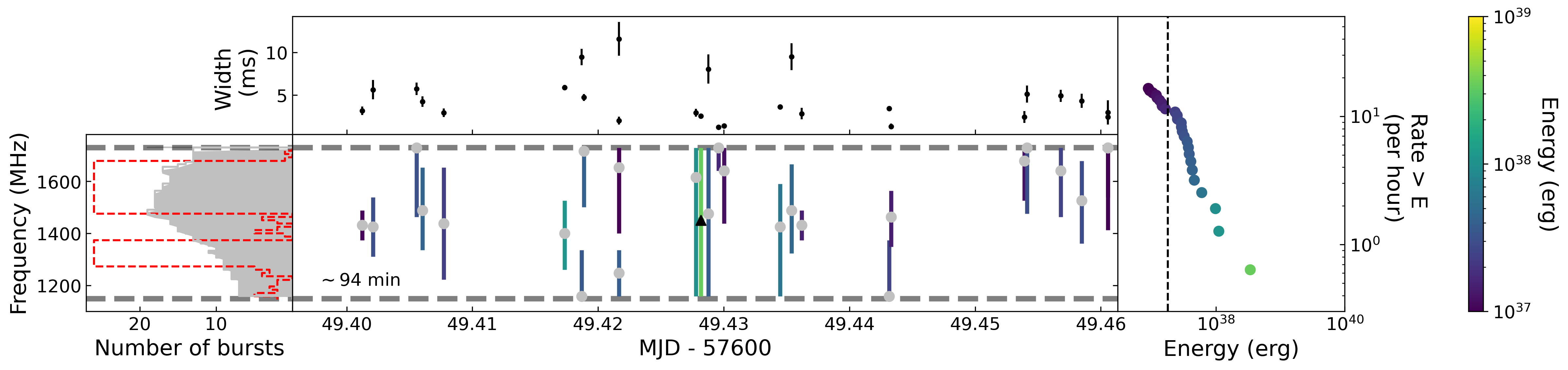}
    \includegraphics[width=1\textwidth]{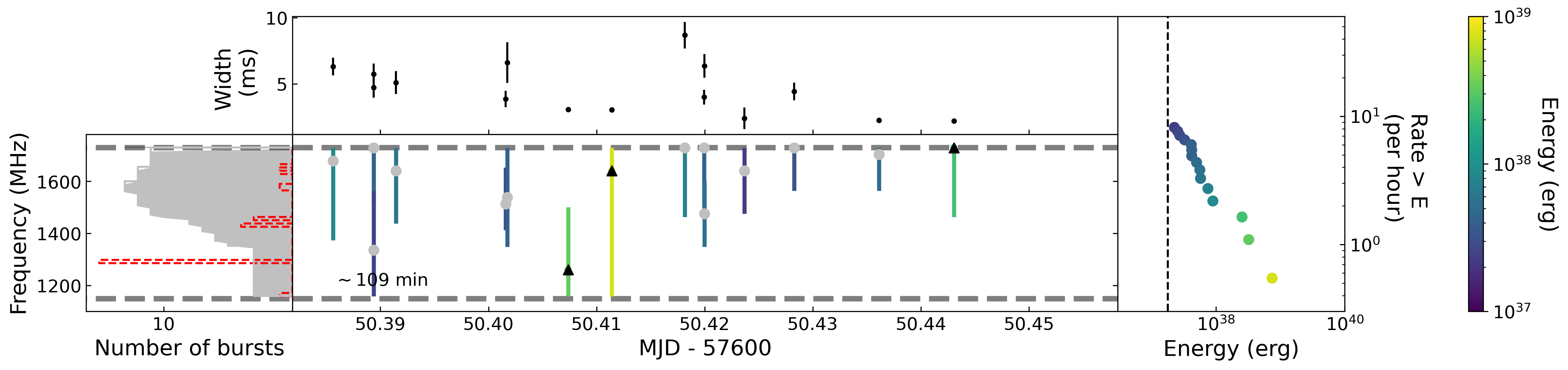}
    \includegraphics[width=1\textwidth]{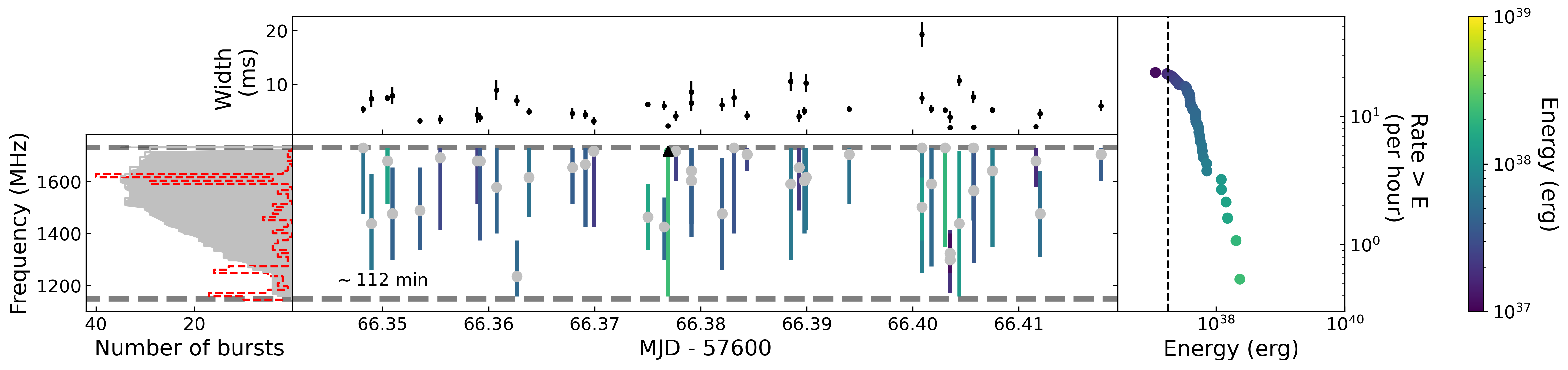}

	\caption{Same as \ref{fig:multiplot1}, but for MJDs 57649, 57650 and 57666.}
    \label{fig:multiplot4}
\end{figure*}

%%%%%%%%%%%%%%%%%%%%%%%%%%%%%%%%%%%%%%%%%%%%%%%%%%

% Don't change these lines
\bsp	% typesetting comment
\label{lastpage}
\end{document}